\documentclass[twocolumn,superscriptaddress,floatfix,preprintnumbers]{revtex4-1}
\usepackage{dcolumn}
\usepackage{amsmath}
\usepackage{graphicx,epsfig,psfrag}
\usepackage{amssymb}
\usepackage{natbib}
\usepackage{url}
\usepackage[breaklinks=true]{hyperref}
\usepackage{mathtools}
\usepackage{subfigure}
\hypersetup{
        colorlinks=true,
}
\usepackage{bookmark}
\usepackage{epic}

\usepackage{txfonts}
\usepackage{color}

\begin{document}
\title{Time evolution of the Kondo resonance in response to a quench}
\author{H. T. M. Nghiem}
\affiliation
{Peter Gr\"{u}nberg Institut and Institute for Advanced Simulation, 
Research Centre J\"ulich, 52425 J\"ulich, Germany}
\affiliation{
Advanced Institute for Science and Technology, Hanoi University of
Science and Technology, 10000 Hanoi, Vietnam}
\author{T. A. Costi}
\affiliation
{Peter Gr\"{u}nberg Institut and Institute for Advanced Simulation, 
Research Centre J\"ulich, 52425 J\"ulich, Germany}
\begin{abstract}
We investigate the time evolution of the Kondo resonance in response to a quench by applying the 
time-dependent numerical renormalization group (TDNRG) approach to the Anderson impurity model in the strong correlation limit. For this purpose, we derive within TDNRG a numerically tractable expression for the retarded two-time nonequilibrium Green function $G(t+t',t)$, and its associated time-dependent spectral function, $A(\omega,t)$, for times $t$ both before and after the quench. Quenches from both mixed valence and Kondo correlated initial states to Kondo correlated final states are considered. For both cases, we find that the Kondo resonance in the zero temperature spectral function, a preformed version of which is evident at very short times $t\to 0^{+}$, only fully develops at very long times $t\gtrsim 1/T_{\rm K}$, where $T_{\rm K}$ is the Kondo temperature of the final state. In contrast, the final state satellite peaks develop on a fast time scale $1/\Gamma$ during the time interval $-1/\Gamma \lesssim t \lesssim +1/\Gamma$, where $\Gamma$ is the hybridization strength. Initial and final state spectral functions are recovered in the limits $t\rightarrow -\infty$ and $t\rightarrow +\infty$, respectively. 
Our formulation of two-time nonequilibrium Green functions within TDNRG provides a first step towards using this method as an impurity solver within nonequilibrium dynamical mean field theory. 
\end{abstract}


\maketitle
{\em Introduction.---}
The nonequilibrium properties of strongly correlated quantum impurity models continue to pose a major theoretical challenge. This contrasts with their equilibrium properties, which are largely well understood \cite{Hewson1993}, or can be investigated within a number of highly accurate methods, such as the numerical renormalization group method (NRG) \cite{Wilson1975,KWW1980a,Bulla2008,Gonzalez-Buxton1998},  the continuous time quantum Monte Carlo (CTQMC) approach \cite{Gull2011b}, the density matrix renormalization group \cite{White1992}, or the Bethe ansatz method \cite{Tsvelick1983b,Andrei2013}. Quantum impurity models far from equilibrium are of direct relevance to several fields of research, including charge transfer effects
in low-energy ion-surface scattering \cite{Brako1981,Kasai1987,Merino1998,Langreth1991,Shao1994a,Shao1994b,Pamperin2015,He2010},
transient and steady state effects in molecular and semiconductor quantum dots \cite{Hershfield1992,Hershfield1993,Meir1993,Bruder1994,Kretinin2011,Kretinin2012,Pletyukhov2012,Scott2013,Nordlander1999,Park2002,Kogan2004,Hemingway2014,Jauho1994,Kennes2012a,Cohen2014a,Schmidt2008,Dorda2014,Rosch2003a,Antipov2016}, and also in the context of dynamical mean field theory (DMFT) of strongly correlated lattice models \cite{Metzner1989,Georges1996,Kotliar2004}, as generalized to nonequilibrium \cite{Schmidt2002,Freericks2006,Aoki2014}. In the latter, further progress hinges on
an accurate non-perturbative solution for the nonequilibrium Green functions of an effective quantum impurity model. Such a solution, beyond allowing time-resolved spectroscopies of correlated lattice systems within DMFT to be addressed \cite{Eckstein2008,Freericks2009,Perfetti2006,Loukakos2007,Iyoda2014}, would 
also be useful in understanding time-resolved scanning tunnelling microscopy of nanoscale systems \cite{Loth2010} and proposed cold atom realizations of Kondo correlated states \cite{Nishida2013,Bauer2013,Nishida2016,Riegger2017}, which could be probed with real-time radio-frequency spectroscopy \cite{Goold2011,Knap2012,Cetina2016}.

In this Letter, we use the time-dependent numerical renormalization group (TDNRG) approach \cite{Anders2005,Anders2006,Anders2008a,Anders2008b,Guettge2013,Nghiem2014a,Nghiem2014b} to calculate the retarded two-time Green function, $G(t_{1}=t+t',t_{2}=t)$, and associated spectral function,
$A(\omega,t)$, of the Anderson impurity model in response to a quench at time $t=0$, and apply this to investigate in detail 
the time evolution of the Kondo resonance. 
This topic has been addressed before within several approaches, including the non-crossing approximation \cite{Nordlander1999,Randi2017}, conserving approximations \cite{Bock2016} and within CTQMC for quantum dots out of equilibrium \cite{Cohen2014a}. 
Related work on the temporal evolution of the spin-spin correlation function in the Kondo model and thermalization in the Anderson impurity model following initial state preparations has also been carried out \cite{Lobaskin2005,Heyl2010}. Formulations of the time-dependent spectral function within TDNRG are also available \cite{Anders2008b,Weymann2015}, but only for positive times.
Here, we derive expressions for the two-time Green function and spectral function $A(\omega,t)$ which are numerically tractable at arbitrary times, including negative times. The main advantages of the TDNRG over other approaches for calculating time-dependent spectral functions
is that it can access arbitrary long times ($t\to   \pm \infty$) and arbitrary low temperatures and frequencies, is non-perturbative  
and calculates spectral functions directly on the real frequency axis. It is therefore well suited for investigating the formation in time of the exponentially narrow and low temperature Kondo resonance \footnote{The use of a discretized Wilson chain within TDNRG results, to a small degree, in incomplete
thermalization at $t\to\infty$ in thermodynamic observables \cite{Anders2006,Rosch2012,Nghiem2014a,Nghiem2016}, 
and in spectral functions \cite{Anders2008b,Weymann2015}, but it suffices for a consistent nonperturbative picture of the overall time evolution of $A(\omega,t)$.}

{\em Model and quenches.---}We consider the time-dependent Anderson impurity model,
$H = \sum_{\sigma}\varepsilon_{d}(t)n_{d\sigma}+U(t)n_{d\uparrow}n_{d\downarrow} + \sum_{k\sigma}\epsilon_{k}c^{\dagger}_{k\sigma}c_{k\sigma}
+\sum_{k\sigma} V(c^{\dagger}_{k\sigma}d_{\sigma}+d^{\dagger}_{\sigma}c_{k\sigma})$,
where $\varepsilon_{d}(t)=\theta(-t)\varepsilon_i+\theta(t)\varepsilon_f$ is the energy of the local level, 
$U(t) =\theta(-t)U_i +\theta(t) U_f$ is the local Coulomb interaction, $\sigma$ labels the spin, $n_{d\sigma}=d_{\sigma}^{\dagger}d_{\sigma}$ is the number operator for local electrons with spin $\sigma$,  and $\varepsilon_{k}$ is the kinetic energy of the conduction electrons with constant density of states $\rho(\omega)=\sum_{k}\delta(\omega-\varepsilon_{k})=1/2D$ with $D=1$ the half-bandwidth. We take
$\Gamma\equiv\pi\rho(0)V^2=0.001$ 
throughout and consider two types of quench [referred to subsequently as quench (A) or quench (B)]: (A), from a symmetric Kondo regime with $\varepsilon_{i}=-15\Gamma, U_{i}=30\Gamma$ and a vanishingly small Kondo scale $T^{i}_{\rm K}=3\times 10^{-8}$ \footnote{We use the Bethe ansatz expression 
$T_{\rm K}=\sqrt{\Gamma U/2}e^{-\pi U/8\Gamma + \pi \Gamma/2U}$ valid in the symmetric Kondo limit $U/\pi\Gamma\gg 1$ \cite{Zlatic1983,Hewson1993}.} to a symmetric Kondo regime with $\varepsilon_{f}=-6\Gamma,U_{f}=12\Gamma$ and a larger Kondo scale $T_{\rm K}=2.5\times 10^{-5}$, and, (B), from a mixed valence regime with $\varepsilon_{i}=-\Gamma, U_{i}=8\Gamma$ to a symmetric Kondo regime with $\varepsilon_{f}=-4\Gamma, U_{f}=8\Gamma$ and a Kondo scale $T_{\rm K}=1.0\times 10^{-4}$. 

{\em Spectral function $A(\omega,t)$.---}
We obtain the time-dependent spectral function via $A(\omega,t)=-\frac{1}{\pi} {\rm Im}[G(\omega+i\eta,t)]$, where  $G(\omega+i\eta,t)$, with infinitesimal $\eta>0$, is the Fourier transform of $G(t+t',t)\equiv -i\theta(t')\langle[d_{\sigma}(t+t'), d_{\sigma}^{\dagger}(t)]_{+}\rangle_{\hat{\rho}}$ with respect to the relative time $t'$ and $\hat{\rho}$ denotes the full density matrix of the initial state \cite{Weichselbaum2007,Peters2006,Costi2010}.
In the notation of Ref.~\onlinecite{Nghiem2014a}, we find for the case of positive times \footnote{See Supplementary Material [URL] for derivations and additional results, including Ref.~\onlinecite{Bulla2001}.}
\begin{align}
&G(\omega+i\eta,t) 
=\sum_{m=m_0}^N\sum_{rsq}^{\notin KK'K''}\rho_{sr}^{i\to f}(m)e^{-i(E_{s}^{m}-E_{r}^{m})t}\nonumber\\
&\times\Big(\frac{B^m_{rq}C^m_{qs}}{\omega+E^m_r-E^m_q+i\eta}
+\frac{C^m_{rq} B^m_{qs}}{\omega+E^m_q-E^m_s+i\eta}\Big),\label{eq:gf-positive-times}
\end{align}
where $B=d_{\sigma}$, $C=d_{\sigma}^{\dagger}$, and $\rho_{sr}^{i\to f}(m)=\sum_{e}{_f}\langle sem|\hat{\rho}|rem\rangle_f$ is the full reduced density matrix projected onto the final states \cite{Nghiem2014a}. A somewhat more complicated expression can be derived for negative times \cite{Note3}. From Eq.~(\ref{eq:gf-positive-times}), we see that the spectral function can be calculated highly efficiently at 
all times and frequencies from a knowledge of $\rho_{sr}^{i\to f}(m)$, the final state matrix elements,
and excitations 
at each shell $m$.
Our expressions for $A(\omega,t)$ in the two time domains $t<0$ and $t>0$ recover the initial and final state spectral functions for $t\to -\infty$ and $t\to +\infty$, respectively and satisfy the spectral sum rule $\int_{-\infty}^{+\infty} d\omega A(\omega,t)=1$  exactly \cite{Note3}.
Below, we shall first focus on positive times, where the main time evolution of the Kondo resonance occurs, then on negative to positive times, showing how the high energy final state features in $A(\omega,t)$ evolve from their initial state counterparts already at negative times.

\begin{figure}[t]
\centering 
\includegraphics[width=1\linewidth]{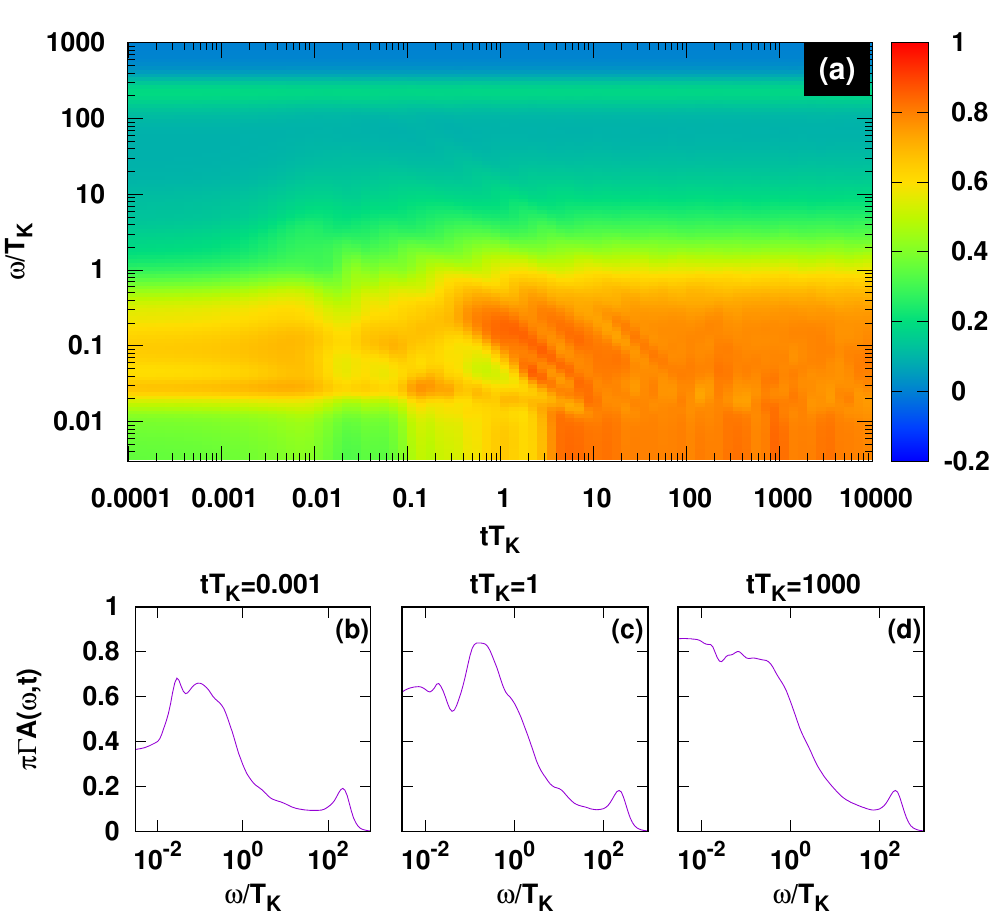}
\caption 
{(a) Time evolution of the normalized spectral function $\pi\Gamma A(\omega>0,t)$ for the symmetric Anderson model at positive times, 
following a quench at $t=0$ specified by  $\varepsilon_{i}=-15\Gamma,U_{i}=30\Gamma$ and $\varepsilon_{f}=-6\Gamma,U_{f}=12\Gamma$ with final state Kondo temperature $T_{\rm K}=2.5\times 10^{-5}$. A structure on the scale of $T_{\rm K}$ evolves into the Kondo resonance at long times $t\gtrsim 1/T_{\rm K}$, while a structure at $\omega = \varepsilon_{f}+U_{f}\approx 240 T_{\rm K}$  with negligible time-dependence corresponds to the 
final state satellite peak. Panels (b)-(d) show the spectral function at times $tT_{\rm K}=0.001, 1$ and $1000$, respectively.
The TDNRG calculations used a discretization parameter $\Lambda=4$, $z$ averaging \cite{Oliveira1994,Campo2005} with $N_z=32$ and a cutoff energy $E_{\rm cut}=24$.
}
\label{fig:symmetry}
\end{figure}

\begin{figure}[t]
\centering 
  \includegraphics[width=1\linewidth]{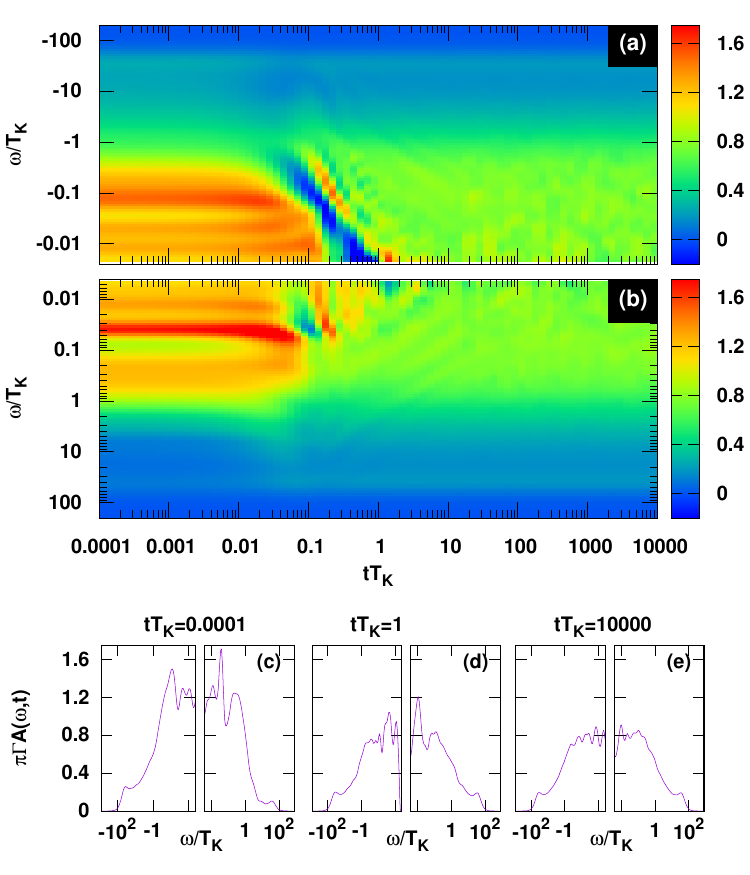}
\caption 
{Time evolution of the normalized spectral function $\pi\Gamma A(\omega,t)$ at positive times,
for, (a), negative, and, (b), positive frequencies, for quench (A) 
from the mixed valence 
to the symmetric Kondo regime
with $T_{\rm K}=1.0\times 10^{-4}$. 
A structure on the scale of $T_{\rm K}$ evolves into the Kondo resonance at long times $t\gtrsim 1/T_{\rm K}$, while structures
at $\omega=\pm \varepsilon_{f}\approx \pm 40T_{\rm K}$, with negligible time-dependence, correspond to 
the final state satellite peaks. Panels (c)-(f) show the  spectral function at times $tT_{\rm K}=0.0001, 1$ and $10000$, respectively.
TDNRG parameters: $\Lambda=4$, $z$ averaging with $N_z=64$ and a cutoff energy $E_{\rm cut}=24$.
}
\label{fig:asymmetry}
\end{figure}

{\em Results for positive times.---}
Consider quench (A), i.e., switching between symmetric Kondo regimes with $T_{\rm K}^{i}\ll T_{\rm K}$. 
Figure~\ref{fig:symmetry}(a) shows the overall time-dependence of the spectral function $A(\omega>0,t>0)=A(-\omega,t>0)$. Two structures, associated with two energy scales, are visible at all times $t>0$ : the satellite peak at $\omega=\varepsilon_{f}+U_{f}\approx 240 T_{\rm K}$ and a structure on the scale of $T_{\rm K}$ around the Fermi level. The former has negligible time-dependence, indicating that the satellite peak in the spectral function has already formed by time $t=0$ (its evolution at negative times from the initial state satellite peak at $\omega=\varepsilon_{i}+U_{i}>\varepsilon_{f}+U_{f}$ is discussed below). In contrast to this, the structure around the Fermi level has significant time-dependence at $t>0$ and evolves into the fully formed final state Kondo resonance only on time scales $t\gtrsim 1/T_{\rm K}$ [Figs.~\ref{fig:symmetry}(c) and \ref{fig:symmetry}(d)] in agreement with Ref.~\onlinecite{Nordlander1999} for the $U=\infty$ Anderson model. 
For $tT_{\rm K}\gg 1$, the height of the Kondo resonance at the Fermi level approaches its unitary value given by the Friedel sum rule $\pi\Gamma  A(\omega=0,t\to\infty)=1$ to within $15\%$ [Fig.~\ref{fig:symmetry}(d)]. The small deviation from the expected value is a result
of incomplete thermalization due to the discretized Wilson chain used within TDNRG \cite{Rosch2012,Weymann2015,Note3}. Consequently,
evaluating $A(\omega,t\to\infty)$ via the self-energy \cite{Bulla1998} does not improve the Friedel sum rule further in this limit \cite{Anders2008b}. In the opposite limit, 
$t\to -\infty$, where thermalization is not an issue, we recover the Friedel sum rule to within $3\%$ (discussed below). The use of a discrete Wilson chain is also the
origin of the small substructures at $|\omega|\lesssim T_{\rm K}$ in Figs.~\ref{fig:symmetry}(b)-\ref{fig:symmetry}(d), effects seen in the time evolution of other quantities, such as the
local occupation, and explained in terms of the discrete Wilson chain \cite{Eidelstein2012}. On shorter time scales, $tT_{\rm K}\lesssim 1$, states in the region $T_{\rm K}^{i}\ll |\omega|<T_{\rm K}$, initially missing [Fig.~\ref{fig:symmetry}(b)], are gradually filled in by a transfer of spectral weight from higher energies [Fig.~\ref{fig:symmetry}(c)] to form the  final state Kondo resonance at long times [Fig.~\ref{fig:symmetry}(d)]. The presence of a structure on the final state Kondo scale $T_{\rm K}$ at short times $t\to 0^{+}$ is understood as follows:
the Fourier transform with respect to $t'=t_{1}-t_{2}$ necessarily convolutes information about the final state at large $t_{1},t_{2}$ into the spectral function at short-times $t$ \cite{Turkowski2005}. Hence, the gross features of the spectral function, even at short times $t\to 0^{+}$, are close to those of the final state spectral function $A(\omega,t\to\infty)$, and far from those of the initial state spectral function. Clear signatures of the latter, such as the much narrower initial state Kondo peak, only appear at negative times. 

Consider now quench (B), in which the system, is switched from the mixed valence to the symmetric Kondo regime. Figures~\ref{fig:asymmetry}(a)-(b) show the overall time-dependence of the spectral function for $\omega<0$ [Figure~\ref{fig:asymmetry}(a)]
and $\omega>0$ [Figure~\ref{fig:asymmetry}(b)]. As for quench (A), two structures associated with two energy scales are again visible at all times $t>0$: the satellite peaks at $\omega=\varepsilon_{f}\approx -40T_{\rm K}$ [Figure~\ref{fig:asymmetry}(a)]  and $\omega=\varepsilon_{f}+U\approx +40 T_{\rm K}$ [Figure~\ref{fig:asymmetry}(b)] and a structure on the scale of $T_{\rm K}$ around the Fermi level [Figs.~\ref{fig:asymmetry}(a) and \ref{fig:asymmetry}(b)]. In contrast to quench (A), the former have some non-negligible time-dependence at short positive times as can be seen in Fig.~\ref{fig:asymmetry}(c) for  $tT_{\rm K}=10^{-4}$ ($t\Gamma=10^{-3}$), where the weight of the satellite peaks has still not equalized.
This asymmetry vanishes on time scales exceeding $1/\Gamma$ [Figs.~\ref{fig:asymmetry}(d) and \ref{fig:asymmetry}(e)  for $tT_{\rm K}=1$ ($t\Gamma=10$)  and $tT_{\rm K}=10^4$ ($t\Gamma=10^3$), respectively]. The low energy structure of width $T_{\rm K}$, initially asymmetric and exceeding the unitary height $1/\pi\Gamma$, 
has significant time-dependence for $t>0$ and evolves into the fully developed Kondo resonance at $t\gtrsim 1/T_{\rm K}$ [Figs.~\ref{fig:asymmetry}(d) and \ref{fig:asymmetry}(e)]. The deviation from the Friedel sum rule  $\pi\Gamma A(\omega=0,t\to\infty)=1$ is comparable to that for quench (A) and reflects the incomplete thermalization due to the discrete Wilson chain used within TDNRG. The discrete Wilson chain also results in the substructures at $|\omega|\lesssim T_{\rm K}$ in Figs.~\ref{fig:asymmetry}(c) and \ref{fig:asymmetry}(d) and in the small remaining asymmetry of the fully developed Kondo resonance in Fig.~\ref{fig:asymmetry}(e).

\begin{figure}[t]
\centering 
\includegraphics[width=1\linewidth]{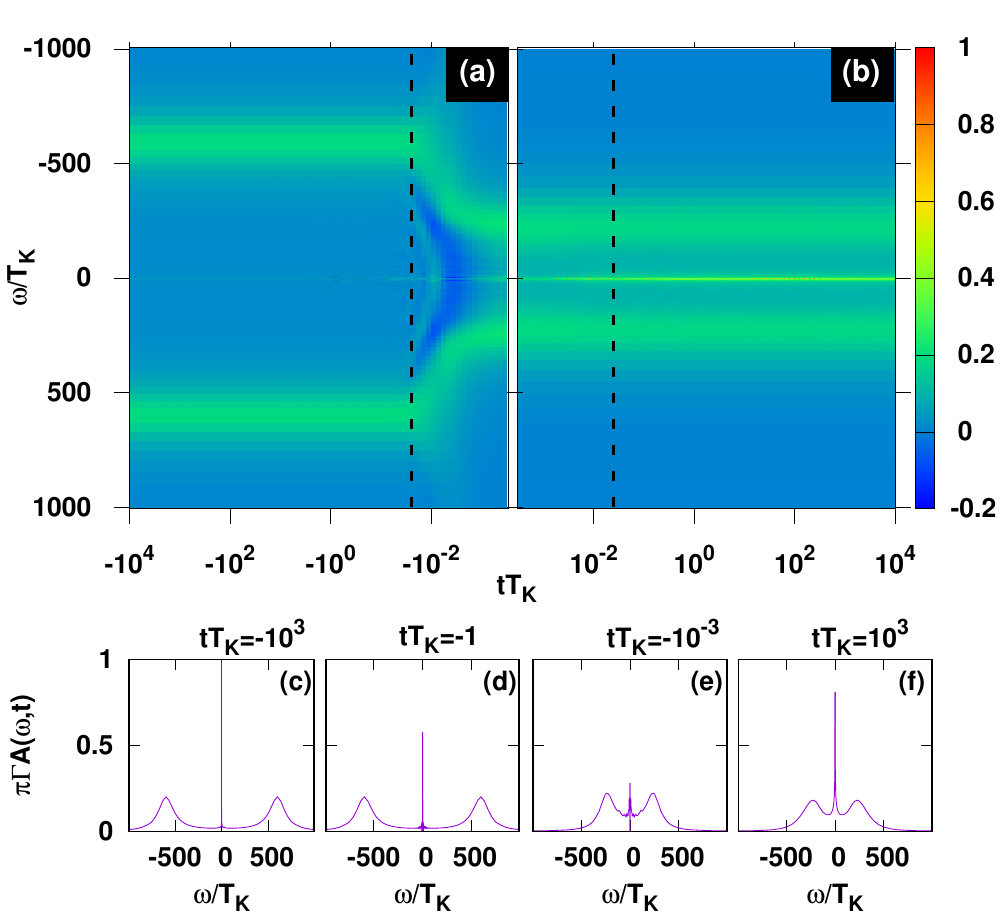}
\caption 
{
$A(\omega,t)$ vs $tT_K$ from,  (a), negative, to, (b), positive times for quench (A), and on a {\em linear} frequency scale. Dashed lines mark $t\Gamma=\pm 1$ ($tT_{\rm K}=\pm 2.5 \times 10^{-2}$). Initial state ($\omega=\pm \varepsilon_{i}=\pm 15\Gamma\approx \pm 600T_{\rm K}$) and final state ($\omega=\pm\varepsilon_{f}=\pm 6\Gamma\approx \pm 240 T_{\rm K}$)  satellite peaks are clearly visible, as are initial and final state Kondo resonances around $\omega=0$. Panels (c)-(e) show the $A(\omega,t)$ 
at times $tT_{\rm K}=-1000, -1,-0.001$ and $+1000$, respectively. TDNRG parameters as in Fig.~\ref{fig:asymmetry}. 
}
\label{fig:symmetry-negative}
\end{figure}

\begin{figure}[h]
\centering 
  \includegraphics[width=1\linewidth]{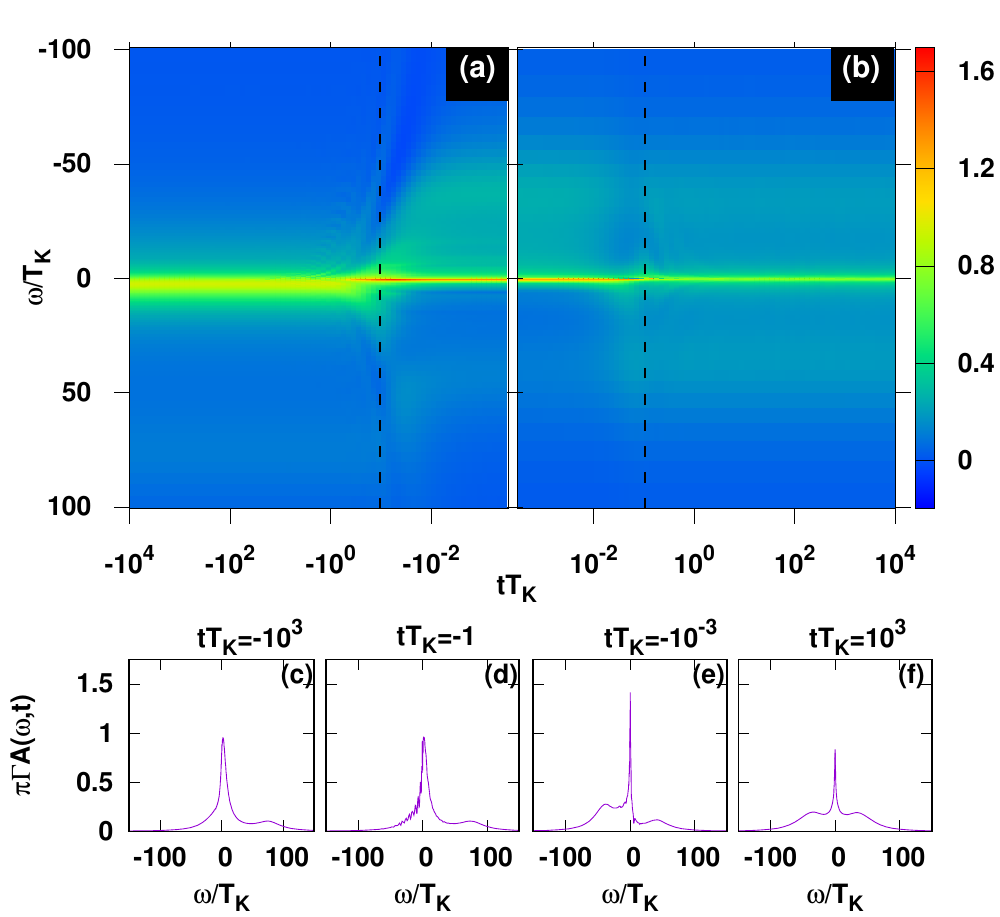}
\caption 
{
$A(\omega,t)$ vs $tT_K$ from, (a), negative, to, (b), positive times and on a {\em linear} frequency scale for quench (B)
(from a mixed valence to a symmetric Kondo regime). Dashed lines mark $t\Gamma=\pm 1$ ($tT_{\rm K}=\pm 10^{-1}$). Panels
(c)-(e) show the spectral function at times $tT_{\rm K}=-1000, -1,-0.001$ and $+1000$, respectively. 
TDNRG parameters as in Fig.~\ref{fig:asymmetry}. 
}
\label{fig:asymmetry-negative}
\end{figure}

{\em From negative to positive times.---}
Figures~\ref{fig:symmetry-negative}(a) and \ref{fig:symmetry-negative}(b) show the overall time-dependence 
of the spectral function for negative and positive times, respectively, for quench (A), on a {\em linear} frequency scale. 
As for positive times [Fig.~\ref{fig:symmetry}(a) and Fig.~\ref{fig:symmetry-negative}(b)], low and high energy structures are visible also for negative times [Fig.~\ref{fig:symmetry-negative}(a)]. 
Moreover, it is clear from Figs.~\ref{fig:symmetry-negative}(a) and \ref{fig:symmetry-negative}(b) that the transition from the initial to the final state spectral function occurs on different time scales for the different structures. Consider
first the high energy structures, which carry essentially all the spectral weight. Initially, these are located at $\omega=\pm\varepsilon_{i}\approx \pm 600 T_{\rm K}$ as is clearly visible in 
Fig.~\ref{fig:symmetry-negative}(a) or in Fig.~\ref{fig:symmetry-negative}(c)  for $tT_{\rm K}=-10^{3}$ ($t\Gamma=-4\times 10^{4}\ll -1$). They cross over to their final state positions at $\omega=\pm\varepsilon_{f}=\pm 240T_{\rm K}$ when $tT_{\rm K}\gtrsim -10^{-2}$ ($t\Gamma\gtrsim -0.4$) [Figs.~\ref{fig:symmetry-negative}(a) and \ref{fig:symmetry-negative}(e)], i.e., on the charge fluctuation time scale $1/\Gamma$. This can also be seen in Figs.~\ref{fig:symmetry-negative}(d) and \ref{fig:symmetry-negative}(e). This large shift in spectral weight
from $\omega=\pm \varepsilon_{i}$ to $\omega=\pm \varepsilon_f$ in the time-range $-10^{-2}\lesssim tT_{\rm K}\lesssim -10^{-3}$ ($-0.4\lesssim t\Gamma\lesssim -0.04$), clearly
seen in Fig.~\ref{fig:symmetry-negative}(a), is accompanied by small regions of negative spectral weight in this transient time range \cite{Note3}. 
This does not violate any exact results for time-dependent, as opposed to steady-state, spectral functions, and is observed in other systems \cite{Dirks2013,Freericks2009b,Jauho1994}.
The spectral sum rule is satisfied analytically exactly at all times and numerically within $1\%$ at all negative times and to higher accuracy 
at positive times for all quench protocols \cite{Note3}. Turning now to the low energy structure, i.e., the Kondo resonance, the use of a linear frequency scale now allows the initial state Kondo resonance at $\omega=0$ to be clearly seen in Fig.~\ref{fig:symmetry-negative}(a) [see also Fig.~\ref{fig:symmetry-negative}(c)].
This structure, of width $T_{\rm K}^{i}\ll T_{\rm K}$ at $t\to -\infty$ and satisfying the Friedel sum rule $\pi\Gamma A(\omega=0,t\to-\infty)=1$, gradually broadens and acquires a width of $T_{\rm K}$ at short negative times \cite{Note3}, and then evolves into the fully developed Kondo resonance on positive time scales $tT_{\rm K}\gtrsim 1$ [Fig.~\ref{fig:symmetry-negative}(e)].

Even more interesting is the negative [Fig.~\ref{fig:asymmetry-negative}(a)] to positive [Fig.~\ref{fig:asymmetry-negative}(b)] time evolution of the spectral function upon quenching from the mixed valence 
to the symmetric Kondo regime [quench (B)]. At large negative times [Fig.~\ref{fig:asymmetry-negative}(c)], one recovers the initial state spectral function of the mixed valence regime (with $\varepsilon_{i}=-\Gamma$) showing a mixed valence resonance, renormalized by many-body effects to lie close to, but just above the Fermi level $\varepsilon_{i}\to \tilde{\varepsilon}_{i}\gtrsim 0$ and satisfying the Friedel sum rule $A(0,t\to -\infty)=\sin^{2}(\pi n_d/2)/\pi\Gamma$ to within $3\%$ \cite{Costi1996b} 
[Figs.~\ref{fig:asymmetry-negative}(a) and \ref{fig:asymmetry-negative}(c), $n_{d}=0.675$]. The upper satellite peak at $\omega = \varepsilon_{i}+U_{i}=7\Gamma\approx 70T_{\rm K}$ is more clearly visible in Fig.~\ref{fig:asymmetry-negative}(c). These peaks give rise to the final state satellite peaks at $\omega=\pm \varepsilon_{f}=\pm 4\Gamma\approx \pm 40 T_{\rm K}$ which start to form already at negative times $tT_{\rm K}\gtrsim -10^{-1}$ ($t\Gamma \gtrsim -1$), i.e., on the charge fluctuation time scale $1/\Gamma$, as for quench (A).  While the positions of these peaks start to shift to their final state values at negative times $tT_{\rm K}\gtrsim -10^{-1}$ ($t\Gamma\gtrsim -1$), their weights
remain disparate [see Fig.~\ref{fig:asymmetry-negative}(e)] and only equalize at $tT_{\rm K}\gtrsim +10^{-1}$ ($t\Gamma \gtrsim +1$) as clearly seen in Fig.~\ref{fig:asymmetry-negative}(b), i.e., the formation of the high energy
final state satellite peaks occurs on a fast time scale $t\approx 1/\Gamma$ in the interval $-1/\Gamma \lesssim t \lesssim +1/\Gamma$ {(dashed lines in Fig.\ref{fig:asymmetry-negative}). 
Going into more details, we see in Figs.~\ref{fig:asymmetry-negative}(a) and \ref{fig:asymmetry-negative}(c)-(e) the deconstruction of the mixed valence resonance in the time range $-1/\Gamma < t <0$. While this resonance carries essentially all the spectral weight at $t\ll -1/\Gamma$, 
weight is gradually transferred to $\omega <0$, with precursor oscillations starting at $tT_{\rm K}=-1$ ($t\Gamma=-10$) [Fig.~\ref{fig:asymmetry-negative}(d)], to form the lower final state satellite peak at $\omega=\varepsilon_{f}$ for $-1/\Gamma < t <0$ [Fig.~\ref{fig:asymmetry-negative}(e)] . Simultaneously, the mixed valence resonance narrows from its original width $\Gamma\approx 10T_{\rm K}$ and shifts towards the Fermi level to form a low energy structure on the scale of $T_{\rm K}$ [Fig.~\ref{fig:asymmetry-negative}(e)]. The latter eventually evolves into the final state Kondo resonance at $tT_{\rm K}\gtrsim 1$. The final state spectral function is recovered in the long-time limit $tT_{\rm K}\gg 1$ [Fig.~\ref{fig:asymmetry-negative}(f)]. 

{\em Conclusions.---}
In summary, we investigated within the TDNRG the time evolution of the spectral function of the Anderson impurity model in the strong correlation limit.
Quenching into a Kondo correlated final state, we showed that the Kondo resonance in the zero temperature spectral function only fully develops at very long times $t\gtrsim 1/T_{\rm K}$, although a preformed version of it is evident even at very short times $t\to 0^{+}$. The latter can be used as a smoking gun signature of the transient build up of the Kondo resonance in future cold atom realizations of the Anderson impurity model \cite{Bauer2013}. The satellite peaks evolve from their initial state values at negative times on a much faster time scale $t\approx  1/\Gamma$ in the time-interval $-1/\Gamma \lesssim t \lesssim 1/\Gamma$.  Our formulation of sum rule conserving two-time nonequilibrium Green functions within TDNRG, including lesser Green functions, and their explicit dependence on both times \cite{Note3}, yields the basic information required for applications to time-dependent quantum transport \cite{Jauho1994} and constitutes a first step towards using TDNRG as an impurity solver within nonequilibrium DMFT \cite{Freericks2006,Gramsch2013,Aoki2014}.

\begin{acknowledgments}
H. T. M. N. thanks Hung T. Dang for fruitful discussions. We acknowledge support from the Deutsche Forschungsgemeinschaft via RTG 1995 and supercomputer support by the John von Neumann institute for Computing (J\"ulich). One of the authors (T. A. C.) acknowledges useful discussions with A. Rosch, J. K. Freericks and the hospitality of the Aspen Center for Physics, supported by the National Science Foundation under grant PHY-1607611, during completion of this work.
\end{acknowledgments}

\pagebreak
\clearpage
\widetext

\onecolumngrid

\begin{center}
\textbf{\large{Supplementary Material on ``
Time evolution of the Kondo resonance in response to a quench ''}}
\end{center}
\setcounter{equation}{0}
\setcounter{figure}{0}
\setcounter{table}{0}
\setcounter{page}{1}
\makeatletter
\renewcommand{\theequation}{S\arabic{equation}}
\renewcommand{\thefigure}{S\arabic{figure}}
\renewcommand{\bibnumfmt}[1]{[S#1]}
\renewcommand{\citenumfont}[1]{S#1}
In this supplementary material, we derive numerically tractable expressions
for the retarded two-time Green function, $G(t+t',t)$, and the associated 
time-dependent spectral function, $A(\omega,t)$, within the single
quench TDNRG for both positive ($t>0$) and negative ($t<0$) times.
For positive times, we compare our numerically tractable expression, obtained within the
full density matrix approach \cite{S_Weichselbaum2007}, to a numerically
more time-consuming expression obtained for positive times only in Ref.~\onlinecite{S_Anders2008b} and
compare spectral densities from the two approaches at selected times
for the Anderson impurity model. 
The $t\to\pm\infty$ and $t\to 0^{\pm}$ limits of $A(\omega,t)$, are
discussed and we prove that our expressions for $A(\omega,t)$ satisfy 
the spectral weight sum rule $\int_{-\infty}^{+\infty}d\omega
A(\omega,t)=1$ exactly analytically. Details of the numerical
evaluation of the spectral functions is given and we show that the
numerical error for the spectral weight sum rule lies below $\approx 1\%$ for all quench
protocols studied. The effect of the discrete Wilson chain on the
Friedel sum rule and the time evolution are discussed. For
completeness, we show results 
for the reverse of quench (B) in the main text,
and make comparisons with non-equilibrium 
non-crossing approximation results at finite low temperature
\cite{S_Nordlander1999} and with a hybridization quench \cite{S_Weymann2015}.
The expresion for the lesser Green function is given and used to calculate the time
dependence of the local occupation number of the Anderson
model. Finally, the explicit dependence of the retarded Green function 
on its two time arguments is illustrated numerically.

\onecolumngrid
\onecolumngrid
\section{Retarded two-time nonequilibrium Green function in TDNRG} 
\label{sec:rgf}
We consider the retarded two-time Green function
$G_{BC}(t+t',t)=-i\theta(t')\operatorname{Tr}\{\hat{\rho}[\hat{B}(t+t'),\hat{C}(t)]_s\}$
for a system undergoing a quantum quench at $t=0$ as described by the 
time-dependent Hamiltonian $H(t)=(1-\theta(t))H_{i}+\theta(t)H_{f}$,
with $\hat{\rho}=e^{-\beta H_{i}}$ the density matrix of the initial state,
represented by the full density matrix in Eq.~(\ref{eq:fdm-initial}) \cite{S_Weichselbaum2007}. Since
$t'>0$, we have two cases to consider, (i), $t>0$, in which case both operators
$B$ and $C$ evolve with respect to $H_{f}$, and, (ii), $t<0$, in
which case, either both operators evolve with respect to $H_{i}$ if
$t<t+t'<0$, or if $t<0<t+t'$ operator $C$ evolves with respect to
$H_i$, while operator $B$ evolves with respect to $H_f$. While the
Fourier transform with respect to the time difference $t'$ yielding
$G(\omega+i\eta,t)$ and hence $A(\omega,t)$ is straightforward in
case (i), in case (ii),  expressions for $G(t+t',t)$ are needed from both time
domains $t<t+t'<0$ and $t<0<t+t'$ in order to construct 
$G(\omega+i\eta,t<0)$ and hence $A(\omega,t<0)$. Depending on the
physical system considered, both positive and negative times may be of
interest. Thus, in problems where the quench represents an initial
state preparation, the main interest is in the evolution at $t>0$
following this preparation \cite{S_Lobaskin2005}. However, if the quench is considered to be
a perturbation applied to the system at time $t=0$, the full
time evolution is of interest. This case is also required for
applications to nonequilibrium dynamical field theory
(DMFT)\cite{S_Freericks2006,S_Turkowski2005,S_Aoki2014}. 
From a theoretical point of view, $t<0$ is also of interest to fully describe the evolution of the spectral
function $A(\omega,t)$ from its initial state value at $t=-\infty$
to its final state value at $t=+\infty$.
\subsection{Positive time-dependence $t>0$}
\label{subsec:positive-times}
We first consider the case $t>0$, treating $t<0$ in the
next subsection. We have for the retarded Green function,
\begin{align}
G_{BC}(t+t',t)&=-i\theta(t')\operatorname{Tr}\{\hat{\rho}[\hat{B}(t+t'),\hat{C}(t)]_s\}\nonumber\\
&=-i\theta(t')\operatorname{Tr}\{\hat{\rho}[e^{iH_f(t+t')}\hat{B}e^{-iH_f(t+t')},e^{iH_ft}\hat{C}e^{-iH_ft}]_s\}\nonumber\\
&=-i\theta(t')\operatorname{Tr}\{e^{-iH_ft}\hat{\rho}e^{iH_ft}[e^{iH_ft'}\hat{B}e^{-iH_ft'},C]_s\}\nonumber\\
&=-i\theta(t')\operatorname{Tr}\{\hat{\rho}(t)[\hat{B}(t'),\hat{C}]_s\}\label{eq:Gre},
\end{align}
where $s=\pm 1$ for fermionic/bosonic Green functions, respectively.
In the notation of Ref.~\onlinecite{S_Nghiem2014a} and following the
approach of Anders \cite{S_Anders2008b}, we have for the first ($BC$) term of the anticommutator with $t>0$
\begin{align}
I_1(t+t',t)=-&i\operatorname{Tr}\{\hat{\rho}(t)\hat{B(t')}\hat{C}\}\nonumber\\
=-&i\sum_{m=m_0}^N\sum_{le}{_f}\langle lem|\hat{\rho}(t)\hat{B}(t')\hat{C}|lem\rangle_f\nonumber\\
=-&i\sum_{m=m_0}^N\sum_{rs}^{\notin KK'}\sum_{ee'}{_f}\langle sem|\hat{\rho}(t)|re'm\rangle_f{_f}\langle re'm|\hat{B}(t')\hat{C}|sem\rangle_f\label{eq:Itt},
\end{align}
where $|lem\rangle, m=m_{0},\dots,N$ is the complete set of eliminated
states in the NRG diagonalization procedure, with $m_{0}$ the first
iteration at which states are eliminated and $N$ is the last NRG
iteration (see Refs.~\cite{S_Anders2006,S_Nghiem2014a} for details).
We evaluate ${_f}\langle re'm|\hat{B}(t')\hat{C}|sem\rangle_f$ by
inserting the decomposition of unity \cite{S_Anders2006} $I=I_m^+ +I_m^-$ between $\hat{B}(t')$ and $\hat{C}$, 
\begin{align}
{_f}\langle re'm|\hat{B}(t')\hat{C}|sem\rangle_f&={_f}\langle re'm|\hat{B}(t')(I_m^+ +I_m^-)\hat{C}|sem\rangle_f\nonumber\\
&=\sum_{qe''}{_f}\langle re'm|\hat{B}(t')|qe''m\rangle_f{_f}\langle qe''m|\hat{C}|sem\rangle_f\nonumber\\
&+\sum_{m'=m_0}^{m-1}\sum_{le''}{_f}\langle re'm|\hat{B}(t')|le''m'\rangle_f{_f}\langle le''m'|\hat{C}|sem\rangle_f\label{eq:BtCf}.
\end{align}
The first term in the above expression is diagonal in the environment
variables $e$, $e'$ , and $q$ runs over all states (kept and
discarded) at the shell $m$. We put $1=1_{m'}^+ +1_{m'}^-$ in the last term to obtain
\begin{align}
&\sum_{m'=m_0}^{m-1}\sum_{le''}{_f}\langle re'm|(1_{m'}^+ +1_{m'}^-)\hat{B}(t')|le''m'\rangle_f{_f}\langle le''m'|\hat{C}(1_{m'}^+ +1_{m'}^-)|sem\rangle_f\nonumber\\
=&\sum_{m'=m_0}^{m-1}\sum_{le''}\sum_{k_1e_1,k_2e_2}{_f}\langle re'm|k_1e_1m'\rangle_f{_f}\langle k_1e_1m'|\hat{B}(t')|le''m'\rangle_f{_f}\langle le''m'|\hat{C}|k_2e_2m'\rangle_f{_f}\langle k_2e_2m'|sem\rangle_f\label{eq:BtCf2}.
\end{align}
Substituting (\ref{eq:BtCf2}) into (\ref{eq:BtCf}) and using the NRG approximation, we have
\begin{align}
{_f}\langle re'm|\hat{B}(t')\hat{C}|sem\rangle_f=&\sum_{q}e^{i(E^m_r-E^m_q)t'}B^m_{rq}\delta_{ee'}C^m_{qs}\nonumber+\\+&\sum_{m'=m_0}^{m-1}\sum_{le''}\sum_{k_1k_2}{_f}\langle re'm|k_1e''m'\rangle_f B^{m'}_{k_1l}e^{i(E^{m'}_{k_1}-E^{m'}_{l})t'} C^{m'}_{lk_2} {_f}\langle k_2e''m'|sem\rangle_f\label{eq:BtCf3}.
\end{align}
Substituting $\quad{_f}\langle
re'm|k_1e''m'\rangle_f=\delta_{e'_me''_m}[A^{\alpha_m\dagger}_{XK}...A^{\alpha_{m'+1}\dagger}_{KK}]_{rk_1}$
and $\quad{_f}\langle
k_2e''m'|sem\rangle_f=\delta_{e''_me_m}[A^{\alpha_{m'+1}}_{KK}...A^{\alpha_m}_{KX'}]_{k_2s}$
into Eq.~(\ref{eq:BtCf3}), results in the following expression for Eq.~(\ref{eq:Itt})
\begin{align}
I_1(t+t',t)=&-i\sum_{m=m_0}^N\sum_{rs}^{\notin KK'}e^{-i(E^m_s-E^m_r)t}\sum_{e}{_f}\langle sem|{\rho}|rem\rangle_f\times\Big\{\sum_{q}e^{i(E^m_r-E^m_q)t'}B^m_{rq}C^m_{qs}\nonumber\\
&+\sum_{m'=m_0}^{m-1}\sum_{lk_1k_2}\sum_{\alpha_m...\alpha_{m'+1}}[A^{\alpha_m\dagger}_{XK}...A^{\alpha_{m'+1}\dagger}_{KK}]_{rk_1} B^{m'}_{k_1l}e^{i(E^{m'}_{k_1}-E^{m'}_{l})t'} C^{m'}_{lk_2} [A^{\alpha_{m'+1}}_{KK}...A^{\alpha_m}_{KX'}]_{k_2s}\Big\},\label{eq:Itt-1}
\end{align}
in which $\rho^{i\to f}_{s,r}(m)=\sum_{e}{_f}\langle sem|{\rho}|rem\rangle_f$
is the projected full reduced density matrix known from
Ref.~\onlinecite{S_Nghiem2014a}. Fourier transforming the above equation
with respect to $t'$ gives
\begin{align}
I_1(\omega+i\eta,t)=&\sum_{m=m_0}^N\sum_{rs}^{\notin KK'}e^{-i(E^m_s-E^m_r)t}\rho^{i\to f}_{s,r}(m)\times\Big\{\sum_{q}\frac{B^m_{rq}C^m_{qs}}{\omega+E^m_r-E^m_q+i\eta}\nonumber\\
&+\sum_{m'=m_0}^{m-1}\sum_{lk_1k_2}\sum_{\alpha_m...\alpha_{m'+1}}[A^{\alpha_m\dagger}_{XK}...A^{\alpha_{m'+1}\dagger}_{KK}]_{rk_1}\frac{B^{m'}_{k_1l}C^{m'}_{lk_2}}{\omega+E^{m'}_{k_1}-E^{m'}_l+i\eta}[A^{\alpha_{m'+1}}_{KK}...A^{\alpha_m}_{KX'}]_{k_2s}\Big\}\label{eq:Iwt}.
\end{align}
with $\eta$ is a positive infinitesimal. Similarly, the second ($CB$) term of
the anticommutator in (\ref{eq:Gre}) gives us
\begin{align}
I_2(\omega+i\eta,t)=&\sum_{m=m_0}^N\sum_{rs}^{\notin KK'}e^{-i(E^m_s-E^m_r)t}\rho^{i\to f}_{s,r}(m)\times\Big\{\sum_{q}\frac{C^m_{rq}B^m_{qs}}{\omega+E^m_q-E^m_s+i\eta}\nonumber\\
&+\sum_{m'=m_0}^{m-1}\sum_{lk_1k_2}\sum_{\alpha_m...\alpha_{m'+1}}[A^{\alpha_m\dagger}_{XK}...A^{\alpha_{m'+1}\dagger}_{KK}]_{rk_1}\frac{C^{m'}_{k_1l}B^{m'}_{lk_2}}{\omega+E^{m'}_{l}-E^{m'}_{k_2}+i\eta}[A^{\alpha_{m'+1}}_{KK}...A^{\alpha_m}_{KX'}]_{k_2s}\Big\}\label{eq:Iwt2}.
\end{align}
Hence, for positive times we obtain
$G(\omega,t)=I_1(\omega,t)+I_2(\omega,t)$. 

In order to calculate the time-dependent spectral function from the above, one can
follow Anders \cite{S_Anders2008b}  by defining the following
time-dependent density matrix
\begin{align}
&\tilde{\rho}^{i\to f}_{k_2k_1}(m',t)=\sum_{m=m'+1}^{N}\sum^{\notin KK'}_{rs}\sum_{\alpha_m...\alpha_{m'+1}}[A^{\alpha_{m'+1}}_{KK}...A^{\alpha_m}_{KX'}]_{k_2s}e^{-i(E^m_s-E^m_r)t}\rho^{i\to f}_{s,r}(m)[A^{\alpha_m\dagger}_{XK}...A^{\alpha_{m'+1}\dagger}_{KK}]_{rk_1}\nonumber\\
&=\begin{dcases}
     0 & \text{if } m'=N;\\
\sum_{\alpha_{m'+1}}\Big\{\sum^{\notin KK'}_{rs}A^{\alpha_{m'+1}}_{k_2s}e^{-i(E^{m'+1}_s-E^{m'+1}_r)t}\rho^{i\to f}_{sr}(m'+1)A^{\alpha_{m'+1}\dagger}_{rk_1}+\sum_{kk'}A^{\alpha_{m'+1}}_{k_2k}\tilde{\rho}^{i\to f}_{kk'}(m'+1,t)A^{\alpha_{m'+1}\dagger}_{k'k_1}\Big\} & \text{otherwise}
   \end{dcases}.\label{eq:recursion}
\end{align}
Then we have the following expression for the Green's function
\begin{align}
G(\omega,t)=&\sum_{m=m_0}^N\sum_{rs}^{\notin KK'}e^{-i(E^m_s-E^m_r)t}\rho^{i\to f}_{s,r}(m)\times\sum_{q}\Big\{\frac{B^m_{rq}C^m_{qs}}{\omega+E^m_r-E^m_q+i\eta}+\frac{C^m_{rq}B^m_{qs}}{\omega+E^m_q-E^m_s+i\eta}\Big\}\nonumber\\
&+\sum_{m'=m_0}^{N-1}\sum_{k_1k_2}\tilde{\rho}^{i\to f}_{k_2k_1}(m',t)\sum_l\Big\{\frac{B^{m'}_{k_1l}C^{m'}_{lk_2}}{\omega+E^{m'}_{k_1}-E^{m'}_l+i\eta}+\frac{C^{m'}_{k_1l}B^{m'}_{lk_2}}{\omega+E^{m'}_{l}-E^{m'}_{k_2}+i\eta}\Big\}, \label{eq:gf-old-positive}
\end{align}
from which the time-dependent spectral function can be
calculated. This expression of Anders \cite{S_Anders2008b}, generalized
here within the full density matrix approach, and hence valid at arbitrary
temperature \cite{S_Nghiem2014a},  requires, however, the time-dependent
reduced density matrix $\tilde{\rho}^{i\to f}_{k_2k_1}(m',t)$ at each
point in time,  and the latter is in turn obtained via the recursion relation
in Eq.~(\ref{eq:recursion}), resulting in a numerically highly time consuming
calculation. This motivated us to develop an alternative and
numerically more tractable expression for the retarded two-time Green function, to which we now turn.

A different and numerically more feasible expression for
$G(\omega,t>0)$ than the expression above, can be obtained if we go back to
Eq.~(\ref{eq:BtCf3}) and substitute this into Eq.~(\ref{eq:Itt}),
\begin{align}
I_1(t+t',t)=-&i\sum_{m=m_0}^N\sum_{rs}^{\notin KK'}\sum_{ee'}{_f}\langle sem|\hat{\rho}(t)|re'm\rangle_f\times\Big(\sum_{q}e^{i(E^m_r-E^m_q)t'}B^m_{rq}\delta_{ee'}C^m_{qs}\nonumber+\\+&\sum_{m'=m_0}^{m-1}\sum_{le''}\sum_{k_1k_2}{_f}\langle re'm|k_1e''m'\rangle_f B^{m'}_{k_1l}e^{i(E^{m'}_{k_1}-E^{m'}_{l})t'} C^{m'}_{lk_2} {_f}\langle k_2e''m'|sem\rangle_f\Big)
\label{eq:Itt2}.
\end{align}
The first term in this expression is simply
\begin{align}
-i\sum_{m=m_0}^N\sum_{rs}^{\notin KK'}\sum_{e}{_f}\langle sem|\hat{\rho}(t)|rem\rangle_f\sum_{q}e^{i(E^m_r-E^m_q)t'}B^m_{rq}C^m_{qs}
\label{eq:Itt2a},
\end{align}
and the second term can be rearranged as follows
\begin{align}
&-i\sum_{m=m_0}^N\sum_{rs}^{\notin KK'}\sum_{ee'}{_f}\langle sem|\hat{\rho}(t)|re'm\rangle_f\sum_{m'=m_0}^{m-1}\sum_{le''}\sum_{k_1k_2}{_f}\langle re'm|k_1e''m'\rangle_f B^{m'}_{k_1l}e^{i(E^{m'}_{k_1}-E^{m'}_{l})t'} C^{m'}_{lk_2} {_f}\langle k_2e''m'|sem\rangle_f\nonumber\\
=&-i\sum_{m'=m_0}^{N-1}\sum_{le''}\sum_{k_1k_2}\sum_{m=m'+1}^N\sum_{rs}^{\notin KK'}\sum_{ee'}{_f}\langle k_2e''m'|sem\rangle_f{_f}\langle sem|\hat{\rho}(t)|re'm\rangle_f{_f}\langle re'm|k_1e''m'\rangle_f B^{m'}_{k_1l}e^{i(E^{m'}_{k_1}-E^{m'}_{l})t'} C^{m'}_{lk_2} \nonumber\\
=&-i\sum_{m'=m_0}^{N-1}\sum_{le''}\sum_{k_1k_2}\sum_{kk'}\sum_{ee'}{_f}\langle k_2e''m'|kem'\rangle_f{_f}\langle kem'|\hat{\rho}(t)|k'e'm'\rangle_f{_f}\langle k'e'm'|k_1e''m'\rangle_f B^{m'}_{k_1l}e^{i(E^{m'}_{k_1}-E^{m'}_{l})t'} C^{m'}_{lk_2}\nonumber\\
=&-i\sum_{m'=m_0}^{N-1}\sum_{le}\sum_{kk'}{_f}\langle kem'|\hat{\rho}(t)|k'em'\rangle_f B^{m'}_{k'l}e^{i(E^{m'}_{k'}-E^{m'}_{l})t'}C^{m'}_{lk} 
\label{eq:Itt2b}.
\end{align}

Combining the above two terms, we have
\begin{align}
I_1(t+t',t)=&-i\sum_{m=m_0}^N\sum_{rs}^{\notin KK'}\sum_{e}{_f}\langle sem|\hat{\rho}(t)|rem\rangle_f\sum_{q}e^{i(E^m_r-E^m_q)t'}B^m_{rq}C^m_{qs}\nonumber\\
&-i\sum_{m=m_0}^{N-1}\sum_{kk'}\sum_{e}{_f}\langle kem|\hat{\rho}(t)|k'em\rangle_f \sum_{l}B^{m}_{k'l}e^{i(E^{m}_{k'}-E^{m}_{l})t'} C^{m}_{lk} \nonumber\\
=&-i\sum_{m=m_0}^N\sum_{rsq}^{\notin KK'K''}\sum_{e}{_f}\langle sem|\hat{\rho}(t)|rem\rangle_f e^{i(E^m_r-E^m_q)t'}B^m_{rq}C^m_{qs}\label{eq:combining},
\end{align}
in which $\sum_{e}{_f}\langle sem|\hat{\rho}(t)|rem\rangle_f=
e^{i(E^m_r-E^m_s)t}\rho^{i\to f}_{sr}(m)$ and all the other matrix
elements are known quantities. Together with the second  term ($CB$) in the anticommutator,
\begin{align}
I_2(t+t',t)=&-i\sum_{m=m_0}^N\sum_{rsq}^{\notin KK'K''}\sum_{e}{_f}\langle sem|\hat{\rho}(t)|rem\rangle_f C^m_{rq}e^{i(E^m_q-E^m_s)t'}B^m_{qs},
\end{align}
we have the retarded Green's function as follows
\begin{align}
G(t+t',t)
=&-i\sum_{m=m_0}^N\sum_{rsq}^{\notin KK'K''}\sum_{e}{_f}\langle sem|\hat{\rho}(t)|rem\rangle_f \Big(e^{i(E^m_r-E^m_q)t'}B^m_{rq}C^m_{qs}+C^m_{rq}e^{i(E^m_q-E^m_s)t'}B^m_{qs}\Big).
\end{align}
This expression is useful for the non-equilibrium DMFT, which
requires the dynamical fields expressed in terms of two-time Green
functions \cite{S_Freericks2006,S_Gramsch2013,S_Aoki2014}. Fourier
transforming with respect to the time difference $t'$ gives
\begin{align}
G(\omega+i\eta,t)
=&\sum_{m=m_0}^N\sum_{rsq}^{\notin KK'K''}\sum_{e}{_f}\langle sem|\hat{\rho}(t)|rem\rangle_f \Big\{\frac{B^m_{rq}C^m_{qs}}{\omega+E^m_r-E^m_q+i\eta}+\frac{C^m_{rq}B^m_{qs}}{\omega+E^m_q-E^m_s+i\eta}\Big\}, \label{eq:gf-new-positive}
\end{align}
which together with  $\sum_{e}{_f}\langle
sem|\hat{\rho}(t)|rem\rangle_f=\rho^{i\to f}_{sr}(m)e^{-i(E_{s}^{m}-E_{r}^{m})t}$
results in a time-dependent spectral function $\displaystyle
A(\omega,t)=-{\rm Im}[G(\omega,t)]/\pi$ that can be evaluated at all
times in a numerically highly efficient manner: only the time-independent
projected density matrix  $\rho^{i\to f}_{sr}(m)$ together with the
NRG excitations and matrix elements are required to evaluate $A(\omega,t)$ at all positive times. While the
above expression looks deceptively similar to the first term in Anders
expression for the positive time Green function in
Eq.~(\ref{eq:gf-old-positive}), this is not the case [note the
different sum $\sum_{rsq}^{\notin KK'K''}$ in
Eq.~(\ref{eq:gf-new-positive}) as compared to the sum
$\sum_{rs}^{\notin KK'}\dots\times \sum_{q\in (K''D)}$ in the first term of
Eq.~(\ref{eq:gf-old-positive})]. It includes also
the second term in Eq.~(\ref{eq:gf-old-positive}), involving $\tilde{\rho}^{i\to
  f}_{k_2k_1}(m',t)$, but within a different approximation that follows from 
the second line of Eq.~(\ref{eq:combining}). In this way, the
recursive evaluation of a reduced density matrix depending explicitly on time is
circumvented in our  expression, making it numerically tractable. A
similar expression to Eq.~(\ref{eq:gf-new-positive}) has been derived
for initial state density matrices corresponding to either pure states, such as
the ground state, or to decoupled initial states
($\Gamma=0$ in the Anderson model) with or without excitations in the
bath and used to study thermalization in the Anderson model following
such an initial state preparation \cite{S_Weymann2015}.

As a check on Eq.~(\ref{eq:gf-new-positive}), we can verify that it 
reduces to the equilibrium retarded Green function in the case that 
$H^f=H^i$ (vanishing quench size). In this limit, using the  definition of the full density matrix of the initial
state\cite{S_Weichselbaum2007}, 
\begin{align}
\hat{\rho} &\equiv \sum_{l'e'm'}{_f}|l'e'm'\rangle_iw_{m'}\frac{e^{-\beta
  E^{m'}_{l'}}}{\tilde{Z}_{m'}}{_i}\langle l'e'm'|,\label{eq:fdm-initial}
\end{align}
and of the reduced full density matrix $R^m_{red}(k,k')$ in
Refs.~\onlinecite{S_Costi2010,S_Nghiem2014a}, we have
\begin{align}
\sum_{e}{_f}\langle
  sem|\hat{\rho}(t)|rem\rangle_f&=\sum_{e}\sum_{l'e'm'}{_f}\langle
                                  sem|e^{-iH^ft}|l'e'm'\rangle_iw_{m'}\frac{e^{-\beta E^{m'}_{l'}}}{\tilde{Z}_{m'}}{_i}\langle l'e'm'|e^{iH^ft}|rem\rangle_f\nonumber\\
&=\sum_{l'}\omega_m\frac{e^{-\beta E^m_{l'}}}{Z_m}\delta_{sl'}\delta_{l'r}+\sum_{kk'}R^m_{red}(k,k')\delta_{sk}\delta_{k'r}. 
\end{align}
Substituting this expression into Eq.~(\ref{eq:gf-new-positive}) for 
$G(\omega,t)$, we obtain the following time-independent expression 
\begin{align}
G(\omega)=&\sum_{m=m_0}^N\sum_{lq}\Big(\omega_m\frac{e^{-\beta E^m_{l}}}{Z_m}\Big)\times\Big(\frac{B^m_{lq}C^m_{ql}}{\omega+E^m_{l}-E^m_q+i\eta}+\frac{C^m_{lq}B^m_{ql}}{\omega+E^m_q-E^m_{l}+i\eta}\Big)\nonumber\\
+&\sum_{m=m_0}^{N-1}\sum_{lkk'} R^{m}_{red}(k,k')\times\Big(\frac{B^{m}_{k'l}C^{m}_{lk}}{\omega+E^{m}_{k'}-E^{m}_l+i\eta}+\frac{C^{m}_{k'l}B^{m}_{lk}}{\omega+E^{m}_l-E^{m}_{k}+i\eta}\Big), 
\label{eq:Gw}
\end{align}
which is identical to the expression in the equilibrium case \cite{S_Weichselbaum2007,S_Costi2010,S_Nghiem2014a}. 

It is interesting to compare the spectral function obtained from our 
Eq.~(\ref{eq:gf-new-positive}) with that of Anders obtained from Eq.~(\ref{eq:gf-old-positive}). 
For this purpose, we consider quench (A) of the main text and three
representative times: $t=0$, $t=1/T_{\rm K}$, and $t\to +\infty$. 
From Figs.~\ref{fig:comp-t}(a) and \ref{fig:comp-t}(c) we see that both 
expressions give the same results at $t=0$ and $t\to +\infty$, while 
at finite times small differences arise for $|\omega|\lesssim T_{\rm 
  K}$ [see Fig.~\ref{fig:comp-t}(b)]. As we discussed in connection 
with Eq.~(\ref{eq:gf-old-positive}) above the main advantage of our new 
expression Eq.~(\ref{eq:gf-new-positive}), is that it can be evaluated numerically 
very efficiently. By contrast, to date, no results at finite times have 
been published using Eq.~(\ref{eq:gf-old-positive}) and only results
for infinite times have been published \cite{S_Anders2008b}.

\begin{figure}[t]
    \includegraphics[width=1.0\textwidth]{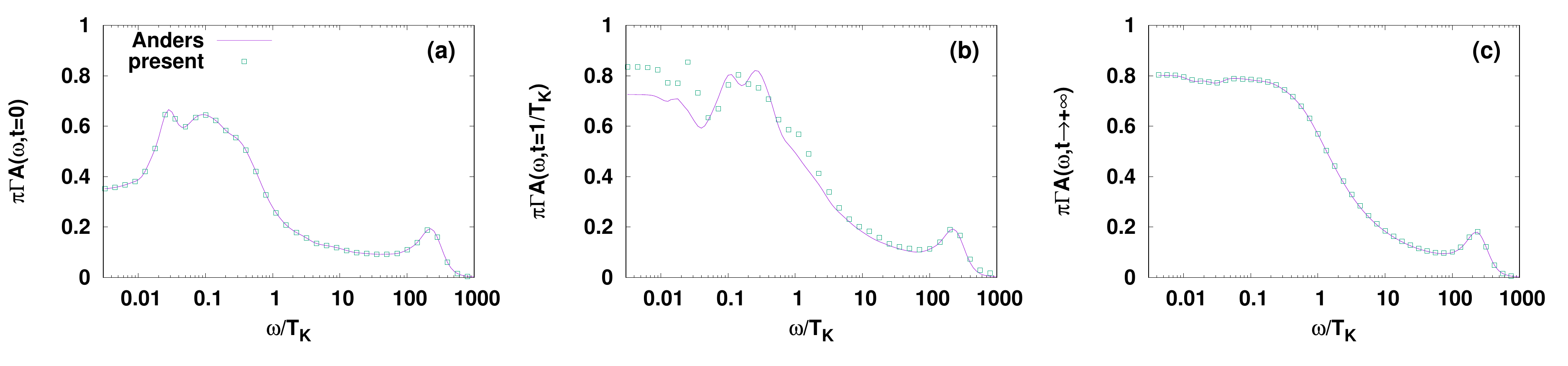}
  \caption{Comparison of the zero temperature spectral functions at different times calculated from 
    our expression [Eq.~(\ref{eq:gf-new-positive})] (symbols) with that from 
    Eq.~(\ref{eq:gf-old-positive}) (solid line). The system is driven 
    from a strongly correlated state to a weaker correlated state as 
    in quench (A) in the main text (where values for quench parameters 
    and the Kondo scale $T_{\rm K}$ can be found). (a) $t=0$, (b) $t=1/T_{\rm K}$, and (c) $t\to +\infty$.}
  \label{fig:comp-t}
\end{figure}

\subsection{Negative time-dependence ($t<0$)}
\label{subsec:negative-times}
At negative times  $t<0$, we can have $t+t'$ before or after the quench. When $t+t'<0$ or $t<-t'$, we have
\begin{align}
G_{BC}(t+t',t)&=-i\theta(t')\operatorname{Tr}\{\hat{\rho}[\hat{B}(t+t'),\hat{C}(t)]_s\}\nonumber\\
&=-i\theta(t')\operatorname{Tr}\{\hat{\rho}[e^{iH_i(t+t')}B e^{-iH_i(t+t')},e^{iH_it}C e^{-iH_it}]_s\}\nonumber\\
&=-i\theta(t')\operatorname{Tr}\{e^{-iH_it}\hat{\rho}e^{iH_it}[e^{iH_it'}B e^{-iH_it'},C ]_s\}\nonumber\\
&=-i\theta(t')\operatorname{Tr}\{\hat{\rho}[e^{iH_it'}B e^{-iH_it'},\hat{C}]_s\},
\end{align}
which is $t$-independent, and just corresponds to the equilibrium propagator of the
initial state Hamiltonian (as long as $t<-t'$). While for $t+t'>0$ or $t>-t'$, we have 
\begin{align}
G_{BC}(t+t',t)&=-i\theta(t')\operatorname{Tr}\{\hat{\rho}[\hat{B}(t+t'),\hat{C}(t)]_s\}\nonumber\\
&=-i\theta(t')\operatorname{Tr}\{\hat{\rho}[e^{iH_f(t+t')}B e^{-iH_f(t+t')},e^{iH_it}C e^{-iH_it}]_s\}\nonumber\\
&=-i\theta(t')\operatorname{Tr}\{e^{-iH_it}\hat{\rho}e^{iH_it}[e^{-iH_it}e^{iH_f(t+t')}B e^{-iH_f(t+t')}e^{iH_it},C ]_s\}\nonumber\\
&=-i\theta(t')\operatorname{Tr}\{\hat{\rho}[e^{-iH_it}e^{iH_f(t+t')}B e^{-iH_f(t+t')}e^{iH_it},\hat{C}]_s\}.
\end{align}
In general, then, we have for the retarded Green function at $t<0$
\begin{equation}
G_{BC}(t+t',t)=\begin{dcases}
     -i\theta(t')\operatorname{Tr}\{\hat{\rho}[e^{iH_it'}B e^{-iH_it'},\hat{C}]_s\} & \text{if } t+t'<0;\\
-i\theta(t')\operatorname{Tr}\{\hat{\rho}[e^{-iH_it}e^{iH_f(t+t')}B e^{-iH_f(t+t')}e^{iH_it},\hat{C}]_s\} &  \text{if } t+t'\geq 0\nonumber,
   \end{dcases}
\end{equation}

For the first part ($BC$ term) of the anticommutator at $t+t'<0$ we have
\begin{align}
I^-_1(t+t',t)&=-i\operatorname{Tr}\{e^{iH_it'}\hat{B} e^{-iH_it'}\hat{C}\hat{\rho}\}\nonumber\\
&=-i\sum_{lem} {_i}\langle lem| e^{iH_it'}\hat{B} e^{-iH_it'}\hat{C}\hat{\rho}|lem\rangle_{i}\nonumber\\
&=-i\sum_{rsem}^{\notin KK'} {_i}\langle rem| e^{iH_it'}\hat{B} e^{-iH_it'}|sem\rangle_{i}\underbrace{{_i}\langle sem|\hat{C}\hat{\rho}|rem\rangle_{i}}_{[C\rho]^m_{sr}}\nonumber\\
&=-i\sum_{rsem}^{\notin KK'} e^{i(E^m_r-E^m_s)t'}B^m_{rs}{[C\rho]^m_{sr}}.\label{eq:BC-term-neg}
\end{align}
Next, we consider $\sum_e[C\rho]^m_{sr}$, which with
Eq.~(\ref{eq:fdm-initial}) becomes
\begin{align}
\sum_e[C\rho]^m_{sr}&=\sum_e{_i}\langle sem|\hat{C}\hat{\rho}|rem\rangle_{i}=\sum_e\sum_{l_1e_1m_1}{_i}\langle sem|\hat{C}|l_1e_1m_1\rangle_i \frac{e^{-\beta E^m_{l_1}}}{\tilde{Z}_{m_1}}w_{m_1}{_i}\langle l_1e_1m_1|rem\rangle_{i}.
\end{align}
In this sum, only the parts with $m_1\geq m$ are finite, and we obtain
\begin{align}
\sum_e[C\rho]^m_{sr}&=\sum_{l}C^m_{sl} \frac{e^{-\beta E^m_l}}{Z_m}w_{m}\delta_{lr}+\sum_{k}C^m_{sk}R^m_{kk'}\delta_{k'r}=\sum_{q}C^m_{sq}\tilde{R}^m_{qr},
\end{align}
with $R_{kk'}^{m}, \tilde{R}^m_{qr}$ as in
Ref.~\onlinecite{S_Nghiem2014a}. Substituting the above into Eq.~(\ref{eq:BC-term-neg}) we obtain
\begin{align}
I^-_1(t+t',t)=-i\sum_{rsm}^{\notin KK'} e^{i(E^m_r-E^m_s)t'}B^m_{rs}\sum_{q}C^m_{sq}\tilde{R}^m_{qr}
\end{align}

Similarly, the first part ($BC$ term) of the anticommutator for $t+t'>0$ is
given by
\begin{align}
&I^+_1(t+t',t)=-i\operatorname{Tr}\{e^{iH_f(t+t')}\hat{B} e^{-iH_f(t+t')}e^{iH_it}\hat{C}\hat{\rho}e^{-iH_it}\}\nonumber\\
&=-i\sum_{lem} {_i}\langle lem| e^{iH_f(t+t')}\hat{B} e^{-iH_f(t+t')}e^{iH_it}\hat{C}\hat{\rho}e^{-iH_it}|lem\rangle_{i}\nonumber\\
&=-i\sum_{rsem}^{\notin KK'} {_i}\langle rem|e^{iH_f(t+t')}\hat{B} e^{-iH_f(t+t')}|sem\rangle_{i}
\underbrace{{_i}\langle sem|e^{iH_it}\hat{C}\hat{\rho}e^{-iH_it}|rem\rangle_{i}}_{[\widetilde{C\rho}(t)]^m_{sr}}\nonumber\\
&=-i\sum_{rsem}^{\notin KK'}\sum_{r_1s_1e_1m_1}^{\notin KK'} {_i}\langle rem|r_1e_1m_1\rangle_{f}{_f}\langle r_1e_1m_1|e^{iH_f(t+t')}\hat{B} e^{-iH_f(t+t')}|s_1e_1m_1\rangle_{f}{_f}\langle s_1e_1m_1|sem\rangle_{i}[\widetilde{C\rho}(t)]^m_{sr},
\end{align}
where $\widetilde{C\rho}(t)\equiv e^{iH_{i}t}C\rho e^{-iH_{i}t}$ (i.e.,
the tilde signifies that the time evolution operators apply to the
composite operator $C\rho$).
We decompose the above sum into three parts corresponding to $m_1>m$, $m_1=m$, and $m_1<m$, then simplify them as follows
\begin{align}
&-i\sum_{rsem}^{\notin KK'}\sum_{kk'}{_i}\langle rem|kem\rangle_{f}{_f}\langle kem|e^{iH_f(t+t')}\hat{B} e^{-iH_f(t+t')}|k'em\rangle_{f}{_f}\langle k'em|sem\rangle_{i}[\widetilde{C\rho}(t)]^m_{sr}\nonumber\\
&-i\sum_{rsem}^{\notin KK'}\sum_{r_1s_1}^{\notin KK'} {_i}\langle rem|r_1em\rangle_{f}{_f}\langle r_1em|e^{iH_f(t+t')}\hat{B} e^{-iH_f(t+t')}|s_1em\rangle_{f}{_f}\langle s_1em|sem\rangle_{i}[\widetilde{C\rho}(t)]^m_{sr}\nonumber\\
&-i\sum_{r_1s_1e_1m_1}^{\notin KK'}\sum_{kk'}{_i}\langle k'e_1m_1|r_1e_1m_1\rangle_{f}{_f}\langle r_1e_1m_1|e^{iH_f(t+t')}\hat{B} e^{-iH_f(t+t')}|s_1e_1m_1\rangle_{f}{_f}\langle s_1e_1m_1|ke_1m_1\rangle_{i}[\widetilde{C\rho}(t)]^{m_1}_{kk'}\nonumber\\
=&-i\sum_{em}\sum_{rsr_1s_1}^{\notin KK'K_1K'_1} {_i}\langle rem|r_1em\rangle_{f}{_f}\langle r_1em|e^{iH_f(t+t')}\hat{B} e^{-iH_f(t+t')}|s_1em\rangle_{f}{_f}\langle s_1em|sem\rangle_{i}[\widetilde{C\rho}(t)]^m_{sr}\nonumber\\
=&-i\sum_{m}\sum_{rsr_1s_1}^{\notin KK'K_1K'_1} S^m_{rr_1}e^{i(E^m_{r_1}-E^m_{s_1})(t+t')}B^m_{r_1s_1} S^m_{s_1s}e^{i(E^m_s-E^m_r)t}\sum_e[C\rho]^m_{sr}.
\end{align}

Substituting $\sum_e[C\rho]^m_{sr}$ into the above equation, we have
\begin{align}
I^+_1(t+t',t)=&-i\sum_{mrsr_1s_1}^{\notin KK'K_1K'_1} S^m_{rr_1}e^{i(E^m_{r_1}-E^m_{s_1})(t+t')}B^m_{r_1s_1} S^m_{s_1s}e^{i(E^m_s-E^m_r)t}\sum_{q}C^m_{sq}\tilde{R}^m_{qr}.
\end{align}
The second part ($CB$ term) of the anticommutators for the two parts
at $t+t'<0$ and $t+t'>0$, denoted by
$I^-_2(t+t',t')$ and $I^+_2(t+t',t')$, respectively, can be derived in a similar
way and combining all terms gives the expression for the two-time
Green function $G(t'+t,t)$ at $t<0$. Fourier transforming with
respect to the time difference $t'$ results in the following
expression for the spectral function at $t<0$
\begin{align}
G&(\omega,t)=\int_{-\infty}^{\infty}dt' e^{i(\omega+i\eta) t'} G(t+t',t)\nonumber\\
=&\int_{0}^{-t}dt' e^{i(\omega+i\eta) t'} (I^-_1(t+t',t)+I^-_2(t+t',t))+\int_{-t}^{\infty}dt' e^{i(\omega+i\eta)) t'} (I^+_1(t+t',t)+I^+_2(t+t',t))\nonumber\\
=&-i\sum_m\Big[\int_{0}^{-t}dt' e^{i(\omega+i\eta) t'} \sum_{rs}^{\notin KK'} e^{i(E^m_r-E^m_s)t'}B^m_{rs}\nonumber\\
&+\int_{-t}^{\infty}dt' e^{i(\omega+i\eta) t'}\sum_{rsr_1s_1}^{\notin KK'K_1K'_1} S^m_{rr_1}e^{i(E^m_{r_1}-E^m_{s_1})(t+t')}B^m_{r_1s_1} S^m_{s_1s}e^{i(E^m_s-E^m_r)t}\Big]\times\sum_{q}(C^m_{sq}\tilde{R}^m_{qr}+\tilde{R}^m_{sq}C^m_{qr})\nonumber\\
=&\sum_m\Big[\sum_{rs}^{\notin KK'} \frac{B^m_{rs}}{\omega+E^m_r-E^m_s+i\eta}(1-e^{-i(\omega+E^m_r-E^m_s+i\eta)t})\nonumber\\
&+\sum_{rsr_1s_1}^{\notin KK'K_1K'_1} S^m_{rr_1}\frac{B^m_{r_1s_1}}{\omega+E^m_{r_1}-E^m_{s_1}+i\eta} S^m_{s_1s}e^{-i(\omega+E^m_r-E^m_s+i\eta)t}\Big]\times\sum_{q}(C^m_{sq}\tilde{R}^m_{qr}+\tilde{R}^m_{sq}C^m_{qr}).\label{eq:gf-negative-times}
\end{align}
From this expression, we easily see that $G(\omega,t\to -\infty)$
recovers the initial state Green function. While from the
starting definition, we have $G(\omega,t\to 0^-)$ equal to
$G(\omega,t\to 0^+)$, in the approximate expressions above this is no
longer guaranteed as a consequence of the NRG approximation
and the different derivations for $t\to 0^{\pm}$. Figures~\ref{fig:continuity}(a)-\ref{fig:continuity}(b) show the spectral functions
at $t\to 0^+$ and $t\to 0^-$ and quantify the size of the
discontinuity at $t=0$. While the spectral functions for $t\to
0^{\pm}$ match to high accuracy at high frequencies $|\omega|\gg T_{\rm
  K}$ for both quench (A) [Fig.~\ref{fig:continuity}(a)] and quench
(B)  [Fig.~\ref{fig:continuity}(b)] of the main text, there is a
mismatch at low frequencies. 

\begin{figure}[t]
  \includegraphics[width=1.0\textwidth]{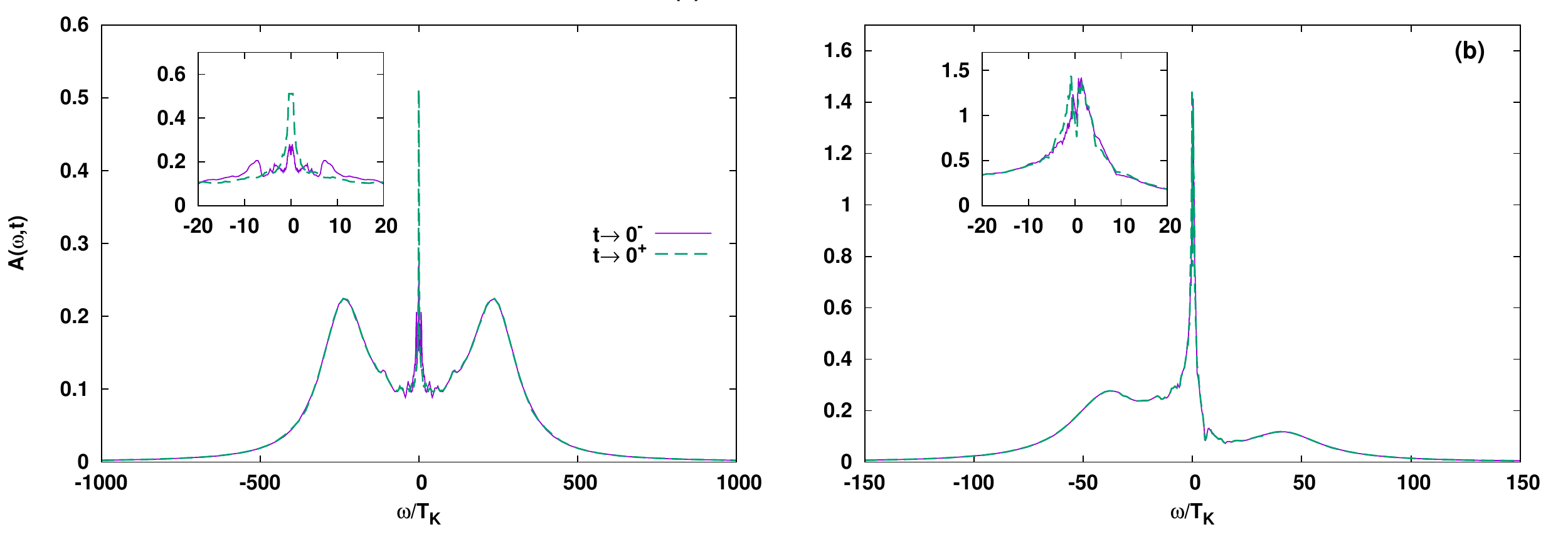}
  \caption{Zero temperature spectral functions at time $t=0^{\pm}$ {\it vs.} 
    scaled frequency $\omega/T_{\rm K}$ for (a) quench (A) of the main 
    text, in which the system is driven from a 
    strongly correlated to a weaker correlated 
    symmetric Kondo regime (with Kondo scale $T_{\rm K}$ as in main 
    text), (b) quench (B) of the main text, in which the system is driven from a 
    an asymmetric mixed valent to a symmetric Kondo regime. Insets show close ups of the low frequency regions.}\label{fig:continuity}
\end{figure}

In Fig.~\ref{fig:log_negative} we show the negative time spectral
function for quench (A) of the main text using logarithmic axes for both time  $t$ (in units
of the initial state Kondo scale $T_{\rm K}^i=1.2\times 10^{-3}T_{\rm K}$) and frequency,
$\omega$ (in units of the final state Kondo scale $T_{\rm K}$). The data is the same as the
positive frequency data in Fig.~3(a) of the main text, but the use of
a logarithmic frequency axis now makes clearer the
statement made there concerning the structure around the Fermi level, 
that ``this structure, of width 
$T_{\rm  K}^{i}\ll T_{\rm K}$  at $t\to -\infty$ and satisfying the Friedel
sum rule  $\pi\Gamma A(\omega=0,t\to-\infty)=1$, gradually broadens 
and acquires a width of $T_{\rm K}$ at short negative times ...''.
In addition, Fig.~\ref{fig:log_negative} shows that the low energy structure on a
scale $T_{\rm K}$ at short negative times is formed by drawing
spectral weight from both the initial state Kondo resonance (diagonal stripes
at $t\gtrsim -1/T_{\rm K}^i$), and also from the satellite peaks
(starting at $t\gtrsim -1/\Gamma$). As for
short positive times [Fig.~1(a) of the main text], the 
``preformed'' Kondo resonance 
at short negative times ($t\to 0^{-}$) is seen to have missing states
in the vicinity ($\omega\ll T_{\rm K}$) of the Fermi level.
Finally, notice that  the relevant time scale for the 
``devolution'' of the initial state Kondo resonance at $t=-\infty$ 
is $1/T_{\rm K}^i=1.2\times 10^{-3}T_{\rm K}$.
\begin{figure}[t]
  \includegraphics[width=0.8\textwidth]{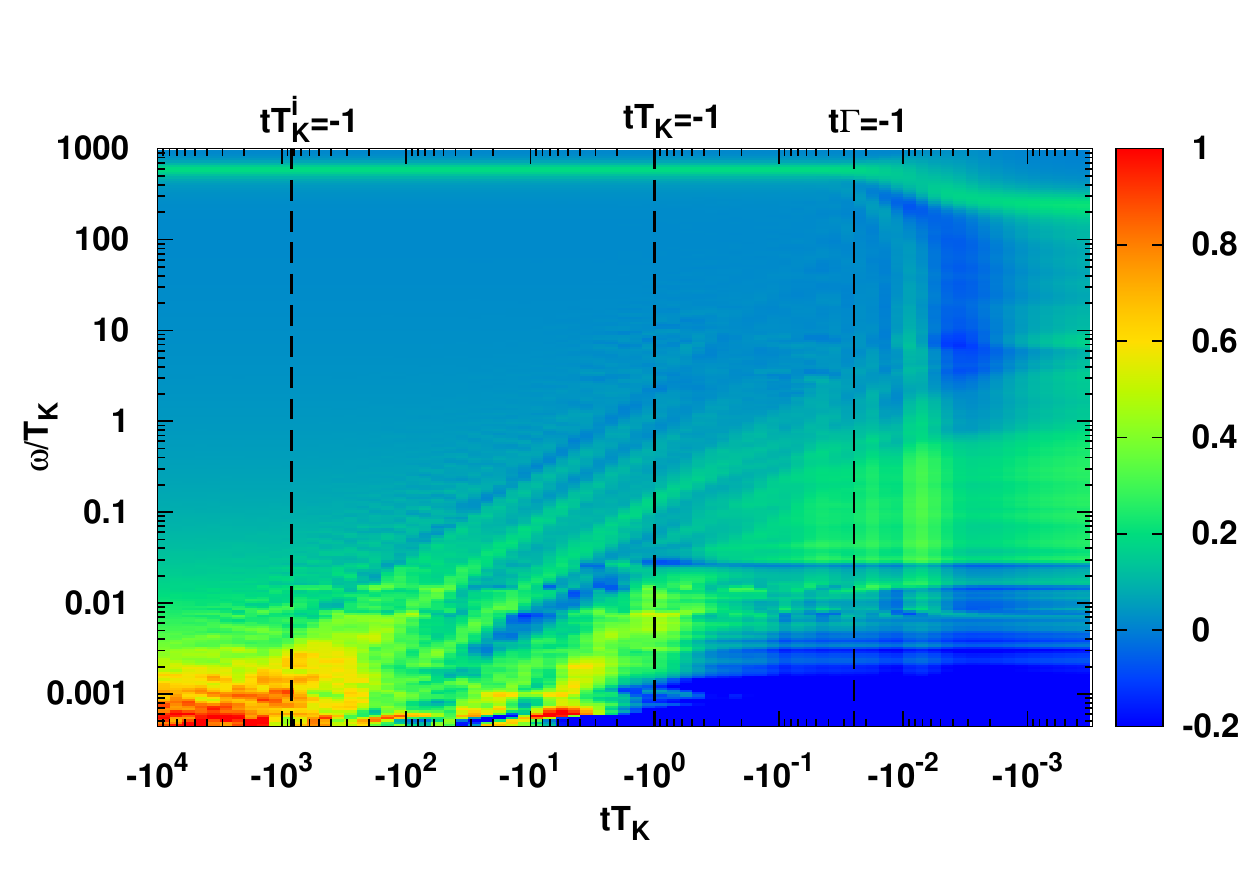}
  \caption{Zero temperature spectral function $A(\omega,t)$ vs 
    $tT_{\rm K}$ at negative times and vs positive frequencies 
    $\omega/T_{\rm K}$ for the symmetric quench (A) of the main text. 
$T_{\rm K}^{i}$ and $T_{\rm K}$ are the Kondo scales of the initial 
and final states respectively. Vertical dashed lines from left to 
right mark $tT_{\rm K}^i=-1$, $tT_{\rm K}=-1$ and 
$t\Gamma=-1$, where $\Gamma$ is the hybridization strength.}
\label{fig:log_negative}
\end{figure}

\subsection{Numerical evaluation of $A(\omega,t)$}
\subsubsection{Positive time spectral function}
We first consider the evaluation of the spectral function for positive
times. From (\ref{eq:gf-new-positive}), we have for
$A(\omega,t)=-\frac{1}{\pi}{\rm Im}G(\omega+i\eta,t)$
\begin{align} 
A(\omega,t)  
=&\sum_{m=m_0}^N\sum_{rsq}^{\notin KK'K''}\rho^{i\to 
   f}_{sr}(m) \cos(E_{sr}^{m}t) 
   \Big\{\delta(\omega-E_{qr}^{m}){B^m_{rq}C^m_{qs}}+\delta(\omega-E_{sq}^m){C^m_{rq}B^m_{qs}}\Big\},\nonumber\\
+& \frac{1}{\pi}\sum_{m=m_0}^N {\rm P.V.}\sum_{rsq}^{\notin KK'K''}\rho^{i\to 
   f}_{sr}(m) \sin(E_{sr}^{m}t) \Big[\frac{B^m_{rq}C^m_{qs}}{(\omega-E_{qr}^{m})}+\frac{C^m_{rq}B^m_{qs}}{(\omega-E_{sq}^m)}\Big], 
 \label{eq:Awt-new-positive}
\end{align} 
where $E_{sr}=E_{s}-E_{r}$. The first contribution,
\begin{align} 
E(\omega,t)  
=&\sum_{m=m_0}^N\sum_{rsq}^{\notin KK'K''}\rho^{i\to 
   f}_{sr}(m) \cos(E_{sr}^{m}t) 
   \Big\{\delta(\omega-E_{qr}^{m}){B^m_{rq}C^m_{qs}}+\delta(\omega-E_{sq}^m){C^m_{rq}B^m_{qs}}\Big\},\label{eq:apos-p1}
\end{align}
is evaluated in the usual way by replacing $\delta(\omega-E)$ by the logarithmic Gaussian 
$\frac{1}{\sqrt{\pi}b|E|}e^{-b^2/4}e^{-(\frac{\ln(|w/E|)}{b})^2}$ and 
summing over the excitations $E$ \cite{S_Bulla2001,S_Bulla2008}. 

In order to evaluate the second 
contribution above, we define the auxiliary function $F''(\omega,t)$ via 
\begin{align}
F''(\omega,t) 
=&\sum_{m=m_0}^N\sum_{rsq}^{\notin KK'K''}\rho^{i\to 
   f}_{sr}(m) \sin(E_{sr}^{m}t) 
   \Big\{\delta(\omega-E_{qr}^{m}){B^m_{rq}C^m_{qs}}+\delta(\omega-E_{sq}^m){C^m_{rq}B^m_{qs}}\Big\}, 
 \label{eq:F-imag}
\end{align}
and evaluate this in the usual way. Taking it's principle value 
integral then gives the second contribution to the spectral function: 
\begin{align}
F'(\omega,t) 
=& -\frac{1}{\pi}{\rm 
   P.V.}\int d\omega'\frac{F''(\omega',t)}{\omega-\omega'}\nonumber\\
=& -\frac{1}{\pi}\sum_{m=m_0}^N {\rm P.V.}\sum_{rsq}^{\notin KK'K''}\rho^{i\to 
   f}_{sr}(m) \sin(E_{sr}^{m}t) \Big[\frac{B^m_{rq}C^m_{qs}}{(\omega-E_{qr}^{m})}+\frac{C^m_{rq}B^m_{qs}}{(\omega-E_{sq}^m)}\Big]. 
\label{eq:F-real}
\end{align}
To summarize,
\begin{align} 
A(\omega,t)  =& E(\omega,t) - F'(\omega,t)\nonumber
\end{align} 
\subsubsection{Negative time spectral function}
We now consider the evaluation of $A(\omega,t)$ for negative
times starting from the expression for the retarded Green function in
Eq.~(\ref{eq:gf-negative-times}). This is a sum of two terms
$G(\omega+i\eta,t)=G_{1}(\omega+i\eta,t)+G_{2}(\omega+i\eta,t)$, with
\begin{align} 
  G_1&(\omega+i\eta,t)=\sum_m\sum_{rs}^{\notin KK'} \frac{B^m_{rs}}{\omega-E^m_{sr}+i\eta}(1-e^{-i(\omega-E^m_{sr}+i\eta)t})\{C\tilde{R}\}_{sr}^m,\label{eq:gf1}\\
  G_2&(\omega+i\eta,t)=\sum_m\sum_{rsr_1s_1}^{\notin KK'K_1K'_1} S^m_{rr_1}\frac{B^m_{r_1s_1}}{\omega-E^m_{s_1r_1}+i\eta} S^m_{s_1s}e^{-i(\omega-E^m_{sr}+i\eta)t}\{C\tilde{R}\}_{sr}^m,\label{eq:gf2}
\end{align} 
where
$\{C\tilde{R}\}_{sr}^m\equiv\sum_{q}(C^m_{sq}\tilde{R}^m_{qr}+\tilde{R}^m_{sq}C^m_{qr})$.  
Correspondingly, the spectral function is also written as a sum of two
parts $A(\omega,t)=A_{1}(\omega,t)+A_2(\omega,t)$.
Consider first $A_{1}(\omega,t)=-\frac{1}{\pi}{\rm Im} G_{1}(\omega,t)$, for finite $0>t>-\infty$, $G_{1}(\omega+i\eta,t)$ is
regular, having no poles on the real axis, then $A_{1}(\omega,t)$ is evaluated directly
\begin{align}
A_{1}&(\omega,t)=\sum_m\sum_{rs}^{\notin KK'}
       \frac{\eta/\pi}{(\omega-E^m_{sr})^2+\eta^2}\{1-e^{\eta 
       t}\cos[(\omega-E^m_{sr})t]\} B^m_{rs}\{C\tilde{R}\}_{sr}^m\nonumber\\
&-\frac{1}{\pi}\sum_m\sum_{rs}^{\notin KK'}
       \frac{(\omega-E^m_{sr})}{(\omega-E^m_{sr})^2+\eta^2}e^{\eta t}\sin[(\omega-E^m_{sr})t] B^m_{rs}\{C\tilde{R}\}_{sr}^m.\label{eq:a1-lorentz}
\end{align}
with a finite $\eta=b|E_{sr}^m|$ and $b\geq 1/N_{z}$ where $N_{z}$ is
the number of bath realizations in the $z$ averaging procedure.


For consistency, we also evaluate $G_{2}(\omega,t)$ in  Eq.~(\ref{eq:gf2})
with the same Lorentzian broadening and thereby obtain
$A_{2}(\omega,t)=-\frac{1}{\pi}{\rm Im} G_{2}(\omega,t)$ and thus 
$A(\omega,t)=A_1(\omega,t)+A_2(\omega,t)$.

\subsection{Spectral sum rule}
\label{subsec:sumrule} 
The spectral weight sum rule $\int_{-\infty}^{+\infty} d\omega A(\omega,t)=1$
with $A(\omega,t)=-\frac{1}{\pi}{\rm Im} G(\omega,t)$ is exactly satisfied for all positive times within our expression
Eq.~(\ref{eq:gf-new-positive}) and the same holds true for Eq.~(\ref{eq:gf-old-positive}) (see Ref.~\onlinecite{S_Anders2008b}).
Here, we prove that it is exactly satisfied also for all negative
times. With $G(\omega,t)$ defined by Eq.~(\ref{eq:gf-negative-times}), we need to evaluate the following terms
\begin{align}
I_1(t)&=-\frac{1}{\pi}\int^{+\infty}_{-\infty}{\rm Im}\Big[\frac{1-e^{-(\omega+E^m_r-E^m_s+i\eta)t}}{\omega+E^m_r-E^m_s+i\eta}\Big]d\omega,\\
I_2(t)&=-\frac{1}{\pi}\int^{+\infty}_{-\infty}{\rm Im}\Big[\frac{e^{-(\omega+E^m_r-E^m_s+i\eta)t}}{\omega+E^m_{r_1}-E^m_{s_1}+i\eta}\Big]d\omega
\end{align}
with $\eta$ a positive infinitesimal. Defining $E^m_r-E^m_s=E^m_{rs}$ and $E^m_{r_1}-E^m_{s_1}=E^m_{r_1s_1}$, we have
\begin{align}
I_1(t)&=\frac{1}{\pi}\int^{+\infty}_{-\infty}\frac{\eta}{(\omega+E^m_{rs})^2+\eta ^2}d\omega
-\frac{1}{\pi}\int^{+\infty}_{-\infty}\frac{e^{\eta t}\eta \cos[(\omega+E^m_{rs})t]}{(\omega+ E^m_{rs})^2+\eta ^2}d\omega
-\frac{1}{\pi}\int^{+\infty}_{-\infty}\frac{e^{\eta t}(\omega+ E^m_{rs}) \sin[(\omega+ E^m_{rs})t]}{(\omega+ E^m_{rs})^2+\eta ^2}d\omega\nonumber\\
&= \frac{1}{\pi} \times \eta \times \frac{\pi}{\eta} -  \frac{1}{\pi} \times e^{\eta t} \eta \times \frac{\pi}{\eta} e^{\eta t} -  \frac{1}{\pi} \times e^{\eta t} \times (-\pi e^{\eta t})\nonumber\\
&= 1 - e^{2\eta t} +   e^{2\eta t} =1,\label{eq:I1}\\
I_2(t)&=\frac{1}{\pi}\int^{+\infty}_{-\infty}\frac{e^{\eta t}\eta \cos[(\omega+ E^m_{rs})t]}{(\omega+ E^m_{r_1s_1})^2+\eta ^2}d\omega
+\frac{1}{\pi}\int^{+\infty}_{-\infty}\frac{e^{\eta t}(\omega+ E^m_{rs}) \sin[(\omega+ E^m_{rs})t]}{(\omega+ E^m_{r_1s_1})^2+\eta ^2}d\omega\nonumber\\
&=\frac{1}{\pi}\int^{+\infty}_{-\infty}\frac{e^{\eta t}\eta \cos[(\omega+ E^m_{r_1s_1})t]\cos[( E^m_{rs}- E^m_{r_1s_1})t]}{(\omega+ E^m_{r_1s_1})^2+\eta ^2}d\omega+\frac{1}{\pi}\int^{+\infty}_{-\infty}\frac{e^{\eta t}(\omega+ E^m_{rs}) \sin[(\omega+ E^m_{r_1s_1})t]\cos[( E^m_{rs}- E^m_{r_1s_1})t]}{(\omega+ E^m_{r_1s_1})^2+\eta ^2}d\omega\nonumber\\
&= \frac{1}{\pi} \times e^{\eta t} \eta \times \frac{\pi}{\eta} e^{\eta t} \cos[( E^m_{rs}- E^m_{r_1s_1})t] +  \frac{1}{\pi} \times e^{\eta t} \times (-\pi e^{\eta t})\cos[( E^m_{rs}- E^m_{r_1s_1})t]\nonumber\\
&= e^{2\eta t}\cos[( E^m_{rs}- E^m_{r_1s_1})t] -   e^{2\eta t} \cos[( E^m_{rs}- E^m_{r_1s_1})t]=0,\label{eq:I2}
\end{align}
where use was made of 
$\int_{-\infty}^{+\infty}\cos(tx)/(x^2+a^2)=\pi e^{ta}/a$ and
$\int_{-\infty}^{+\infty}x\sin(tx)/(x^2+a^2)=-\pi e^{ta}$ ($t<0$).
Using Eqs.~(\ref{eq:I1}-\ref{eq:I2}) and Eq.~(\ref{eq:gf-negative-times}), we have 
$$-\frac{1}{\pi}\int_{-\infty}^{+\infty} d\omega\; {\rm Im}[G(\omega,t)]=\sum_m\sum_{rs}^{\notin KK'}B^m_{rs}\sum_{q}(C^m_{sq}\tilde{R}^m_{qr}+\tilde{R}^m_{sq}C^m_{qr})=1.$$
Therefore the sum rule is proved at $t<0$. The sum rule also holds for
$t=0^{-}$, as can be seen by noting that the last integrals
contributing to $I_1(t)$ and $I_2(t)$ return $0$ in this case.

The numerically evaluated negative-time spectral function also satisfies this sum rule to within $\approx 1\%$ for all quench 
protocols and all negative times, as shown in
Fig.~\ref{fig:sumrule-error}(a). In Fig.~\ref{fig:sumrule-error}(b) we show
$w_{-}(t)=\int_{-\infty}^{+\infty}d\omega F(\omega,t)$ with
$F(\omega,t)=A(\omega,t)$ if $A(\omega,t)<0$ and $F(\omega,t)=0$ if
$A(\omega,t)>0$, i.e., the contribution to the total weight coming from
regions of negative spectral density. Regions of negative spectral
weight appear for transient times in many other systems
\cite{S_Dirks2013,S_Jauho1994,S_Freericks2009b}, while in steady state limits
$t\to\pm\infty$ the spectral function is positive definite within canonical
density matrix approaches. We see that both $w_{-}(t)$ and
the error in the sum rule are both largest in the time region
$t\Gamma\gtrsim -1$ where the major part of the spectral weight
(located in the satellite peaks) is being rearranged 
from $\omega=\varepsilon_i, \varepsilon_i+U_i$ to $\omega=\varepsilon_f,
\varepsilon_f + U_f$. The maximum size of $w_{-}(t)$ correlates with the quench size
$\Delta\epsilon_d=|\epsilon_f-\epsilon_i|$, and reaches up to $15\%$
for the largest quench (A) in Fig.~\ref{fig:sumrule-error}(b).
\begin{figure}[t]
  \includegraphics[width=1.0\textwidth]{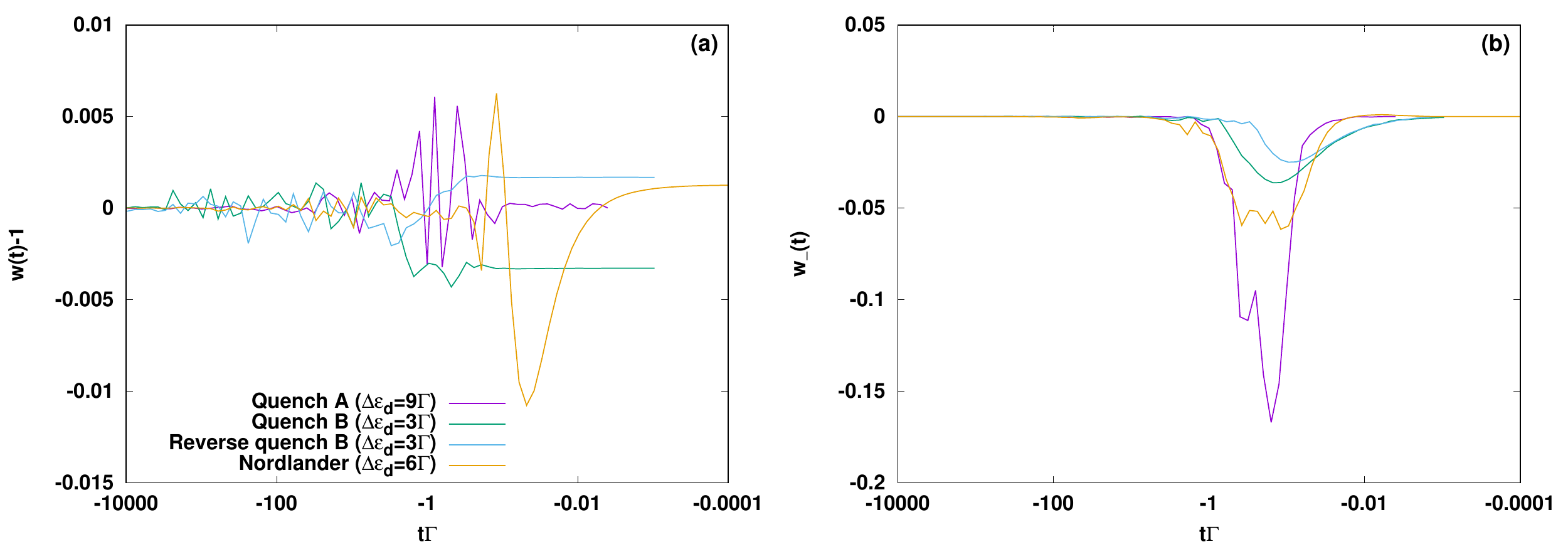}
  \caption{(a) Absolute error in the spectral weight sum rule at negative times, 
    $w(t)-1$ with $w(t)=\int_{-\infty}^{+\infty}d\omega A(\omega,t)$, for the cases of
    quench (A) (protocol 1), quench (B) (protocol 2), reverse of 
    quench (B) (protocol 3) and the quench in Nordlander {\it et al.} in 
    Ref.~\onlinecite{S_Nordlander1999} (protocol 4). 
    (b) The corresponding contribution, $w_{-}(t)$, of negative spectral weight relative to 
    the total weight ($=1$) as a function of time $t<0$ for the four quench 
    protocols. The maximum value of 
    $w_{-}(t)$, close to $t\Gamma \gtrsim -1$, correlates with 
    the    quench size as measured by 
    $\Delta\epsilon_d=|\epsilon_f-\epsilon_i| =
    9\Gamma,3\Gamma,3\Gamma$ and $6\Gamma$ for the four quenches respectively. 
}\label{fig:sumrule-error}
\end{figure}
\subsection{Friedel sum rule, thermalization and discretization effects}
\label{subsec:FSR}
The TDNRG replaces the continuum conduction electron bath
$H=\sum_{k\sigma}\epsilon_{k}c^{\dagger}_{k\sigma}c_{k\sigma}$ in the
Anderson model by a
logarithmically discretized bath $\epsilon_{k}\to \pm \Lambda^{-n}$ whose tight binding representation in
energy space (the so called Wilson chain) is given by
$H=\sum_{n=o,\sigma}^{N}t_{n}(f_{n\sigma}^{\dagger}f_{n+1\sigma}+H.c.)$
with $t_{n}\approx \Lambda^{-(n-1)/2}$ for $n\gg 1$ and where
$\Lambda>1$ is the discretization parameter. The continuum limit
corresponds to $\Lambda\to 1^{+}$, which is not possible to
take numerically within the iterative NRG diagonalization scheme due
to the increasing slow convergence in this limit. The effect of using such a
discrete Wilson chain on the time evolution of physical quantities
in response to a quench is twofold: (i), incomplete thermalization at
long times after the quench due to the fact that a Wilson chain (even
in the limit $N\to\infty$) cannot act as a proper heat bath \cite{S_Rosch2012} (see also
Refs.~\onlinecite{S_Anders2006,S_Nghiem2014a,S_Nghiem2014b,S_Weymann2015}) 
and ,(ii),  additional real features appear in the time evolution,
due to the logarithmic discretization of the bath. We address these
two effects in turn.

Incomplete thermalization in the long time limit is reflected in deviations of observables from
their expected values in the final state. These deviations can be
reduced by decreasing $\Lambda$, as shown in
Ref.~\onlinecite{S_Nghiem2014a} for the case of the local occupation
$n_{d}(t)$ in the Anderson model, where $n_d(t\to+\infty)$ was found
to approach its expected value more closely upon decreasing $\Lambda$ towards $1$. In the present
context of spectral functions, incomplete thermalization is reflected
in a $\approx 15\%$ deviation of 
$\pi\Gamma A(\omega=0,t\to\infty)$ from the continuum result
$\sin^{2}(\pi n_d/2)$,  where $n_d$ is the occupation number in the final state. 
However, this deviation is not an error of the TDNRG but is expected 
because of (i). The Friedel sum rule (FSR) only holds for equilibrium states. 
In the infinite past, where thermalization issues play no role, and 
one achieves the equilibrium state at $t=-\infty$, the FSR is satisfied 
to within a $1-3\%$ for all quenches 
studied, which is comparable to the accuracy achievable in equilibrium NRG
calculations \cite{S_Costi1996b} (the spectra for large negative $t$ in 
Fig.3(c) and Fig.4(c) of the main text).  This demonstrates that the $15\%$ 
deviation in the value of  $\pi\Gamma A(\omega=0,t\to\infty)$ is not 
an  error of the TDNRG  but is the correct result for the Wilson chain used. 

We note, furthermore, that other approaches to time dependent spectral densities,
such as the non-crossing approximation \cite{S_Nordlander1999} suffer,
even for equilibrium spectral densities, from much larger errors in
the Friedel sum rule (see Table I in Ref.~\onlinecite{S_Costi1996b}),
and  still other methods \cite{S_Bock2016} can result in errors 
which  exceed $50\%$ despite the use of a continuum bath. In the
latter approaches the deviations from the Friedel sum rule represent
real errors in the underlying methods, whereas in the TDNRG, the
deviation observed is that expected from using a logarithmically discretized chain.

\begin{figure}[t]
  \includegraphics[width=1.0\textwidth]{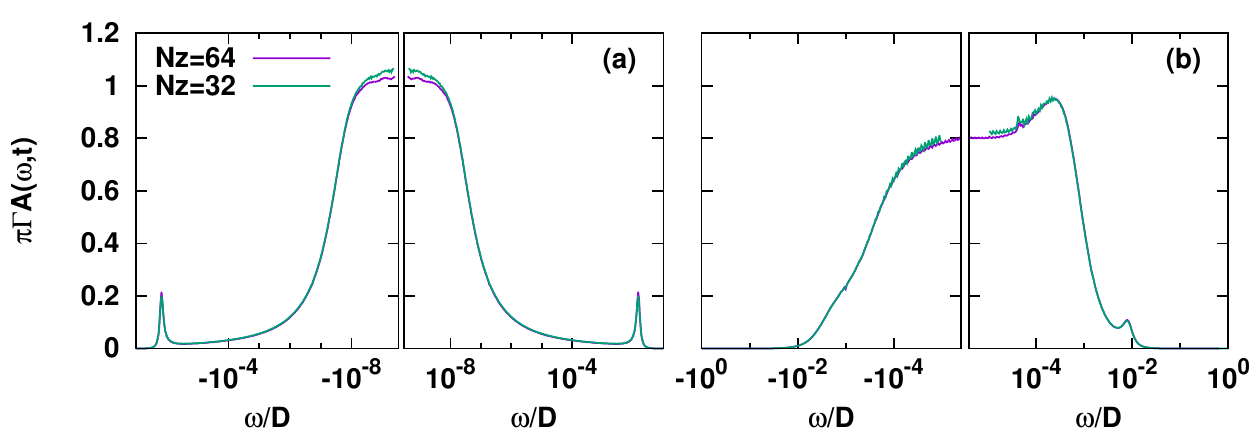}
  \caption{The spectral function in the distant past $A(\omega,t\to  
    -\infty)$, before the quench, for  ,(a), the symmetric quench (A), and, 
    (b), the quench from the mixed valence to symmetric Kondo regime  
    [quench (B)] of the main text. The logarithmic scale shows that  
    substructures associated with reflections are absent at  
    $|\omega|\lesssim T_{\rm K}$, in contrast to the presence of such  
    structures at long positive times. The spectra were evaluated  
    using Eq.~(\ref{eq:a1-lorentz}) with a broadening  
    $\eta=b|E_{sr}^{m}|$ and fixed $b=1/32$. The tiny 
    remnant oscillations, related to the underlying  
    discrete spectrum, are seen to be eliminated by increasing $N_z$. 
}\label{fig:inf_negative}
\end{figure}
 The second effect of using a Wilson chain is that additional
features appear in the time evolution of physical observables that
would be absent for a continuum bath. Examples are the small oscillations 
seen in Figs~.1 and 2 of the main text at low energies $|\omega|<T_{\rm K}$.
Physically, these oscillations, or "substructures" result from the highly
nonequilibrium situation created by the quench: following the quench,
the local change in energy has to be transported by electrons
propagating  outwards in the process of thermalization. These electrons are
reflected off the different sites ($n=0,1,...,N$) of the inhomogeneous
Wilson chain (with hoppings $t_n\sim \Lambda^{-(n-1)/2}$) arriving
back at the impurity site where they interfere at specific times to give
additional features in the time evolution of local quantities (such as
in $A(\omega,t)$). This was originally explained in great detail by
Eidelstein {\it et al.} in Ref.~\onlinecite{S_Eidelstein2012} for observables
like the occupation number $n_{d}(t)$.  These authors also showed a
comparison between $n_d(t)$ using TDNRG and $n_d(t)$ using exact
diagonalization for a noninteracting resonant level model, both
calculated using the same Wilson chain. The additional features in the time
evolution of $n_d(t)$, not present in the continuum model, were
identified as real  effects and not as being due to errors or
unphysical features of the TDNRG method. Thus, the features seen 
 in Figs.1 and 2 of the main text at low energies $|\omega|<T_{\rm K}$
 have their origin in the logarithmically discretized bath. 
In the remote past, before the quench is able to act, such features are expected to
be absent. In support of this interpretation, we therefore show on a
logarithmic scale the spectral function in the distant past
$A(\omega,t=-\infty)$  in Figs.~\ref{fig:inf_negative}(a)-\ref{fig:inf_negative}b, which
indeed show the absence of all ``substructures''. This is to be
compared with the presence of such "substructures" at finite and long
positive times [Figs.~1(b)-1(d) and Fig.~2(c)-2(e) in the main text]. 
Finally, we mention that since the limit $\Lambda\to 1^{+}$ is not
feasible within NRG, a simpler approach that is useful to obtain
results closer to those of the continuum limit is the averaging
of time-dependent quantities over $N_z\gg 1$ realizations of the bath 
\cite{S_Anders2006}.

\section{Additional results}
\label{sec:additional-results}
Our derivation for the non-equilibrium two-time retarded Green
function at both positive and negative times can be used to derive
expressions for the other commonly used Green functions in many-body
theory, including lesser (and greater) Green functions, and can be used 
for arbitrary quenches. In the next
subsections, we show the results for the cases of the reverse of
  quench (B) in the main text, i.e. quenching from a symmetric Kondo
  into a mixed valence regime, a finite temperature quench as in
  Nordlander {\it et al.} \cite{S_Nordlander1999}, and a hybridization quench in which
  the coupling $\Gamma$ is switched on at time $t=0$. We also 
show results for the lesser Green function, which together with
the retarded Green function constitute the basic ingredients for many
applications, e.g., to transient and non-equilibrium transport through correlated
quantum dots \cite{S_Hershfield1992,S_Meir1993,S_Jauho1994}. In
Sec.~\ref{subsec:gf-real-time}, we show the explicit two-time
dependence of the retarded Green function, a basic ingredient in
nonequilibrium DMFT applications \cite{S_Freericks2006}.

\subsection{Symmetric Kondo to mixed valence [reverse of quench (B)]}
In this subsection, we show the time-dependent spectral functions 
for the case of the reverse of quench (B) in the main text. i.e., from
the symmetric Kondo to the mixed valence regime. 
We see in  Fig. \ref{fig:asym_time_rev}  how the spectral function evolves
from the spectral function of the initial state at $t\to -\infty$
[Fig.~\ref{fig:asym_time_rev}(c) for $tT_{\rm K}=-10^3$ ($t\Gamma=-10^4$)],
with the Friedel sum rule satisfied to within a few $\%$ in this
limit, to  its value in the long-time limit with a mixed valence peak
close to the Fermi level and a satellite peak at
$\omega=\varepsilon_{f}+U$ above the Fermi level 
[Fig.~\ref{fig:asym_time_rev}(f) for $tT_{\rm K}=+10^3$ ($t\Gamma=+10^4$)]. 
Similar to the other cases in the main text, the initial state satellite peaks at
$\omega=\pm \varepsilon_{i}$ rapidly relocate at $t=-1/\Gamma$ [dashed
line in Fig.~\ref{fig:asym_time_rev}(a)] with spectral weight being
shifted to form the upper satellite peak and the mixed valence
resonance, a process which essentially has completed by $t=+1/\Gamma$ 
[dashed line in Fig.~\ref{fig:asym_time_rev}(b)]. The weights of these
peaks weights are also close to those of the final state for
$t\Gamma\gtrsim +1$. The central peak which represents the Kondo resonance at $t\to
-\infty$ also varies strongly at $t\gtrsim -1/\Gamma$, and evolves into the
mixed valence peak by time $t\gtrsim +1/\Gamma$. 
\begin{figure}[t]
  \includegraphics[width=1.0\textwidth]{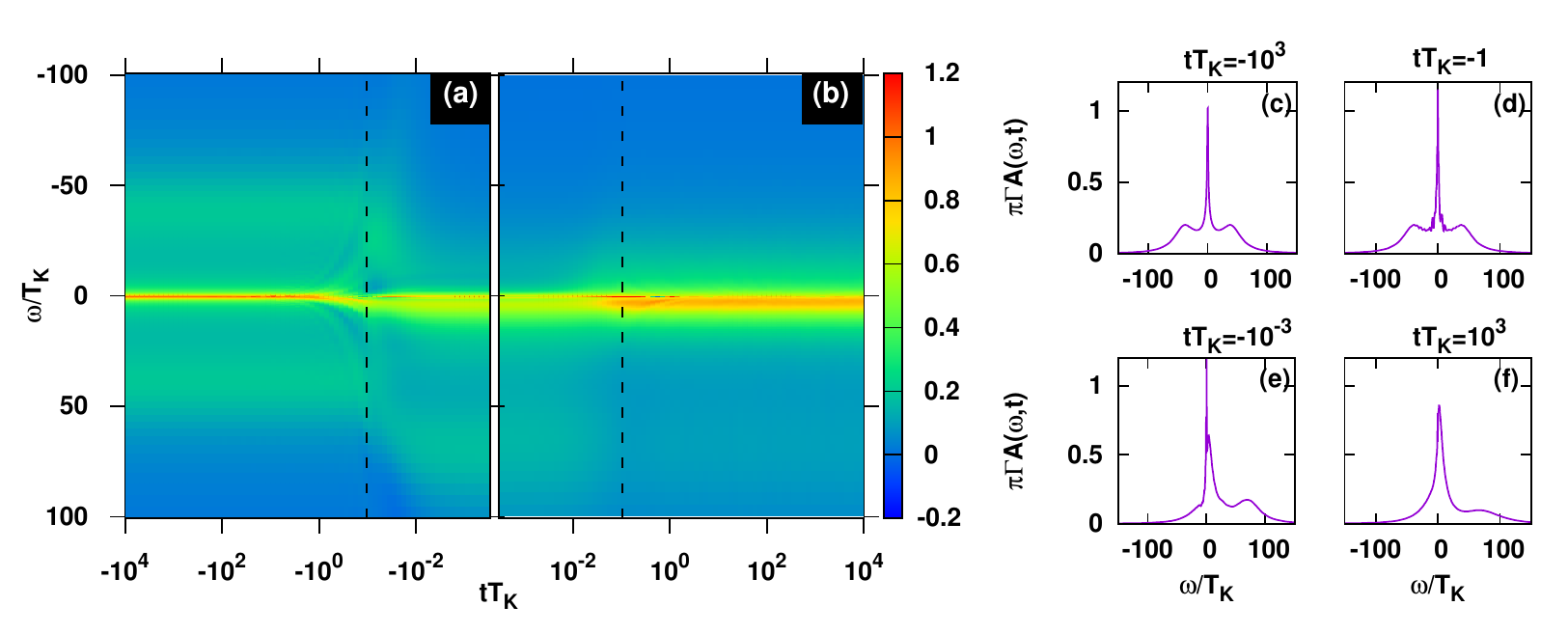}
  \caption{Evolution of the zero temperature spectral function $A(\omega,t)$ for, (a),
    negative, and, (b), positive  times for the reverse quench (B) in
    the main text, and on a {\em linear} frequency scale. 
Spectral functions at representative times are shown in  (c) $tT_K=-10^{3}$, (d) $tT_K=-10^{0}$, (e) $tT_K=-10^{-3}$, and (f) $tT_K=10^{3}$. 
}\label{fig:asym_time_rev}
\end{figure}

\begin{figure}[t]
  \includegraphics[width=1.0\textwidth]{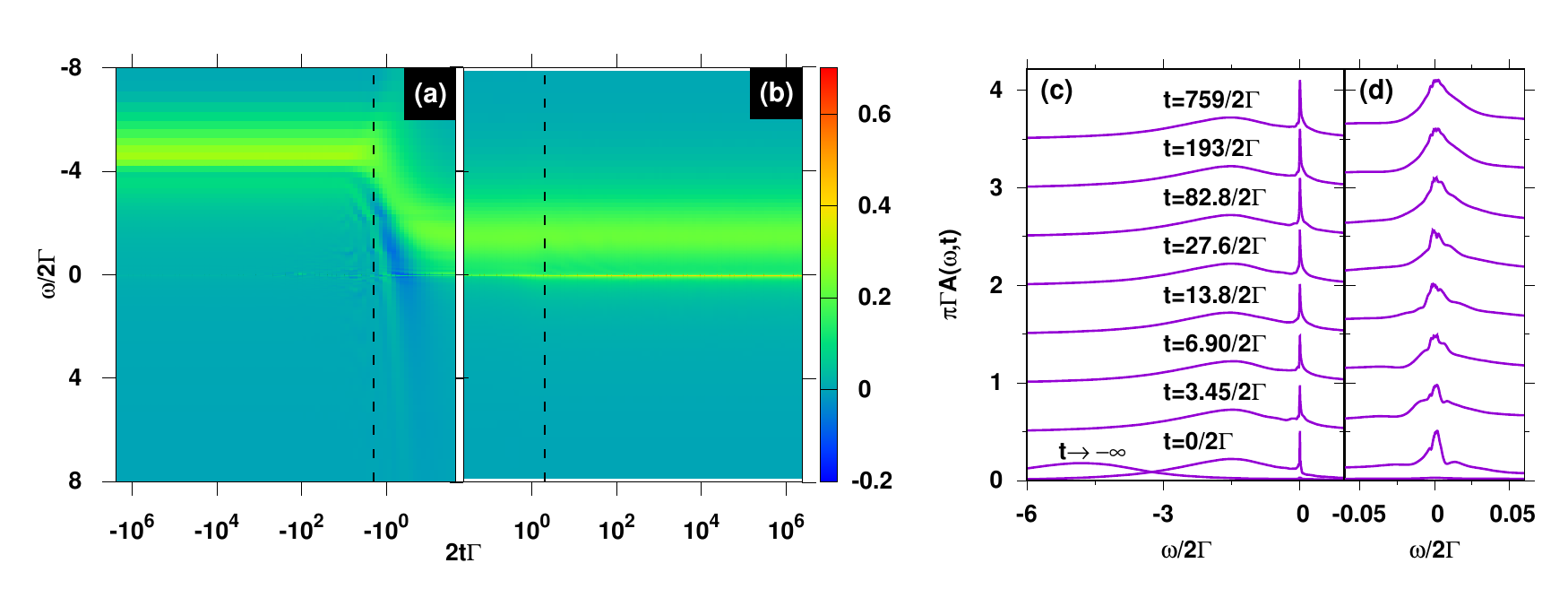}
  \caption{Evolution of the (finite temperature) spectral function $A(\omega,t)$ from, 
    (a), negative, to, (b), positive times for the same quench as in 
    Nordlander {\it et al.} \cite{S_Nordlander1999}, and on a {\em linear}
    frequency scale. Initial state ($\omega=\varepsilon_{i}=
    -10\Gamma$) and final state ($\omega=\varepsilon_{f}=-4\Gamma$) 
    satellite peaks are clearly visible (since $U=\infty$ the upper 
    satellite peaks are absent). We show frequency (time) in units of 
    $2\Gamma$ ($1/2\Gamma$) to ease comparison with results in 
    Ref.~\onlinecite{S_Nordlander1999}, which used a hybridization 
    $\Gamma_{\rm dot}=2\Gamma$ as unit. The initial state Kondo resonance of 
    width $T_{\rm K}^{i}\approx 10^{-7}$ is strongly suppressed by a 
    finite temperature $T=2.5\times 10^{-3}\gg T_{\rm K}^{i}$. The 
    final state Kondo temperature $T_{\rm K}\approx 1.8\times 
    10^{-3}\approx T$. Spectral functions at selected positive times 
    as in Ref.~\onlinecite{S_Nordlander1999} are shown in (c) over a 
    wide frequency, and in (d) over a smaller frequency range around 
    the Fermi level. 
}\label{fig:compare2nca}
\end{figure}
\subsection{Nordlander quench (finite temperature)}
\label{subsec:Nordlander}
In Ref.~\onlinecite{S_Nordlander1999} a quench is made within the 
$U=\infty$ Anderson model via a level shift $\varepsilon_{d}(t) =
\varepsilon_{i}\theta(-t) + \varepsilon_{f}\theta(t)$ with 
$\epsilon_{i}=-10\Gamma$ ($T_{\rm K}^{i}\approx 10^{-7}$) and 
$\epsilon_{f}=-4\Gamma$ ($T_{\rm K}\approx 1.8\times 
    10^{-3}$). This corresponds to a quench from one asymmetric Kondo 
    regime to another with disparate Kondo scales; in contrast we 
    previously investigated quenches in which one of the states was in 
    a symmetric Kondo regime whereas the other was in an mixed valence regime. The quench in 
    Ref.~\onlinecite{S_Nordlander1999} also differs from those studied so far 
    since it is at a finite temperature $T=2.5\times 10^{-3}$ such that  
$T_{\rm K}^{i}\ll T \approx T_{\rm K}$. Thus, initially the Kondo resonance 
is strongly temperature suppressed whereas in the final state it is only 
moderately suppressed by temperature. This 
quench can therefore serve to illustrate the application 
of our TDNRG formalism for time dependent spectral functions to finite temperatures. 

 In figure \ref{fig:compare2nca}, we show the time-dependent spectral 
function from negative to positive times. The calculations were 
carried out for $U\gg D$ to simulate the $U=\infty$ case. We therefore 
observe only the satellite peak below the Fermi level in the negative 
frequency range, both in the initial and final states. 
Similar to the other calculations, the satellite peak rapidly relocates at 
$t\approx -1/\Gamma$ from $\varepsilon_i$ to $\varepsilon_f$ as shown in Fig.~\ref{fig:compare2nca} (a). 
At the same time, the spectral function develops small regions of 
negative spectral weight, with the total sum-rule remaining satisfied 
to within $1\%$ as shown in Fig.~\ref{fig:sumrule-error} (a). 
The central peak at $\omega=0$ is absent at $t\to -\infty$
since the calculation is at finite temperature $T\gg T_{\rm K}^i$. 
Since the temperature $T\approx T_{\rm K}$ is finite and comparable to
the final state Kondo scale, the Kondo resonance does not fully
develop at long times [Fig.~\ref{fig:compare2nca}(b) and \ref{fig:compare2nca}(c)] with
$\pi\Gamma A(\omega=0,t\to\infty)$ reaching only about $59\%$ of its $T=0$ value. 
This is better seen in Fig.~\ref{fig:compare2nca}(d), which shows a
close up of the low frequency region around the Fermi level.
Nevertheless, despite the finite temperature, one sees the build up of
the Kondo resonance at $t\gtrsim 1/T_{\rm K}$.

\subsection{Hybridization quench}
\label{subsec:HybridizationQuench}
In this subsection, we show the time-dependent spectral functions for
the case of a hybridization quench as in
Ref. \onlinecite{S_Weymann2015}, where the hybridization between the
impurity and the conduction electrons, initially turned off at $t<0$,
is suddenly  turned on at $t=0$. 
In figure \ref{fig:compare2Weymann} (a)-(b), we see that for this
quench also, low and high energy features are present at all
times. The high energy features correspond to the final state satellite peaks at
$\varepsilon_f$ and $\varepsilon_f+U$, whereas the low energy feature
of width on the scale of the final state Kondo temperature $T_{\rm K}$
represents the Kondo resonance. While the former have little
temperature dependence at all $t>0$, as in  Weymann {\it et al.}
\cite{S_Weymann2015}, the latter has significant time dependence,
developing fully only at $tT_{\rm K}\gtrsim 1$
[Fig.~\ref{fig:compare2Weymann}(a)] with weight drawn in from higher
energies in the process. Notice that this
low energy peak appears even at $t=0$, which is different from Weymann
{\it et al},\cite{S_Weymann2015} since the broadening parameter is set to be
time-independent in our calculation, while it is time-dependent (and
large of order $\Gamma$ at $t=0$) in Weymann {\it et al.} \cite{S_Weymann2015} . 
While the strong time dependence of the Kondo resonance can be
seen on a logarithmic frequency scale from  Fig.~\ref{fig:compare2Weymann}(a),
it is barely discernible on the linear frequency scale of Fig.~\ref{fig:compare2Weymann}(b).

\begin{figure}[t]
  \includegraphics[width=1.0\textwidth]{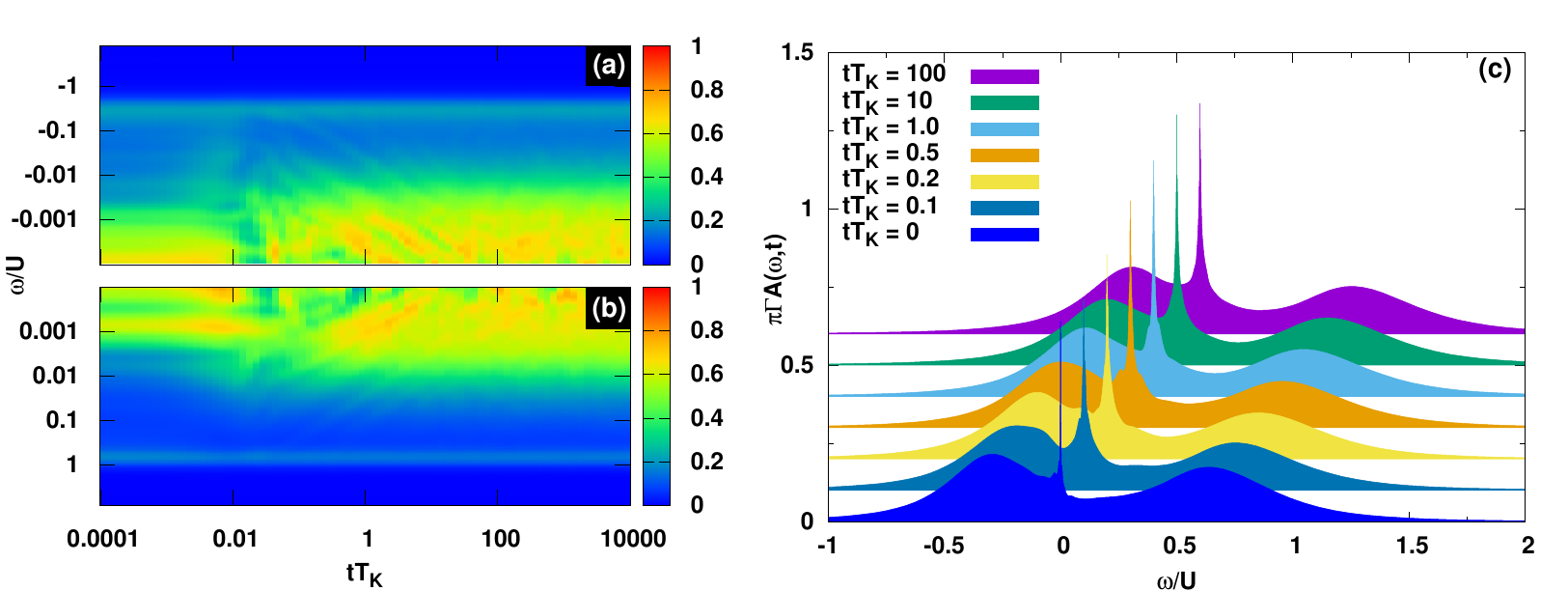}
  \caption{Evolution of the spectral function $A(\omega,t)$ at
    positive times for the same quench as in Weymann {\it et al.} in
    Ref.~\onlinecite{S_Weymann2015}, and on a {\em logarithmic}
    frequency scale, for, (a), negative frequencies,  and, (b),
    positive frequencies. (c) Spectral functions at selected finite positive
    times on  a {\em linear} frequency scale, and $0.1$ vertical
    offsets for clarity. The final state parameters are given by
    $\Gamma=0.01$, $U=12\Gamma$, $\varepsilon_d=-4\Gamma$. The final
    state Kondo scale is $3.7\times 10^{-4}$. The initial
    state parameters are the same, except that $\Gamma=0$.
}\label{fig:compare2Weymann}
\end{figure}

\subsection{Lesser Green functions}
\label{subsec:lesser-gf}
 We consider explicitly the lesser Green function for the local level in
the Anderson impurity model, defined by
\begin{align}
G^<(t+t',t)=i\langle d^\dagger_{\sigma}(t+t')d_{\sigma}(t)\rangle.
\end{align}
For equal times ($t'=0$),
\begin{align}
G^<(t,t)=i\langle d^\dagger_{\sigma}(t)d_{\sigma}(t)\rangle=i\langle n_{d\sigma}(t)\rangle,
\end{align}
i.e., $\text{Im}[G^<(t,t)]=n_{d\sigma}(t)$, so the lesser Green function at equal times
gives the  time evolution of the local occupation number. Following
the derivation for the retarded Green function in Sec.~\ref{sec:rgf},
we similarly obtain the following expression for the lesser Green function 
\begin{align}
G^<(t+t',t)
=&i\sum_{m=m_0}^N\sum_{rsq}^{\notin KK'K''}\sum_{e}{_f}\langle sem|\hat{\rho}(t)|rem\rangle_f e^{i(E^m_r-E^m_q)t'}B^m_{rq}C^m_{qs}\nonumber\\
G^<(t,t)
=&i\sum_{m=m_0}^N\sum_{rsq}^{\notin KK'K''}\sum_{e}{_f}\langle sem|\hat{\rho}(t)|rem\rangle_f B^m_{rq}C^m_{qs},
\end{align}
with $B\equiv d^\dagger_{\sigma}$ and $C\equiv d_{\sigma}$. The
time evolution of the  occupation number calculated from this
expression by setting $t'=0^{+}$ can be compared with that calculated
directly from the thermodynamic observable $n_{d\sigma}(t)$
\cite{S_Nghiem2014a}. The two results, shown in
Fig.~\ref{fig:compare_static_dynamic}, match perfectly at short times
and differ slightly on longer time scales ($t\Gamma\gtrsim 1$).  This
small difference arises because the NRG approximation enters
differently in the expressions for thermodynamic and dynamic quantities.
\begin{figure}[t]
    \includegraphics[width=0.5\textwidth]{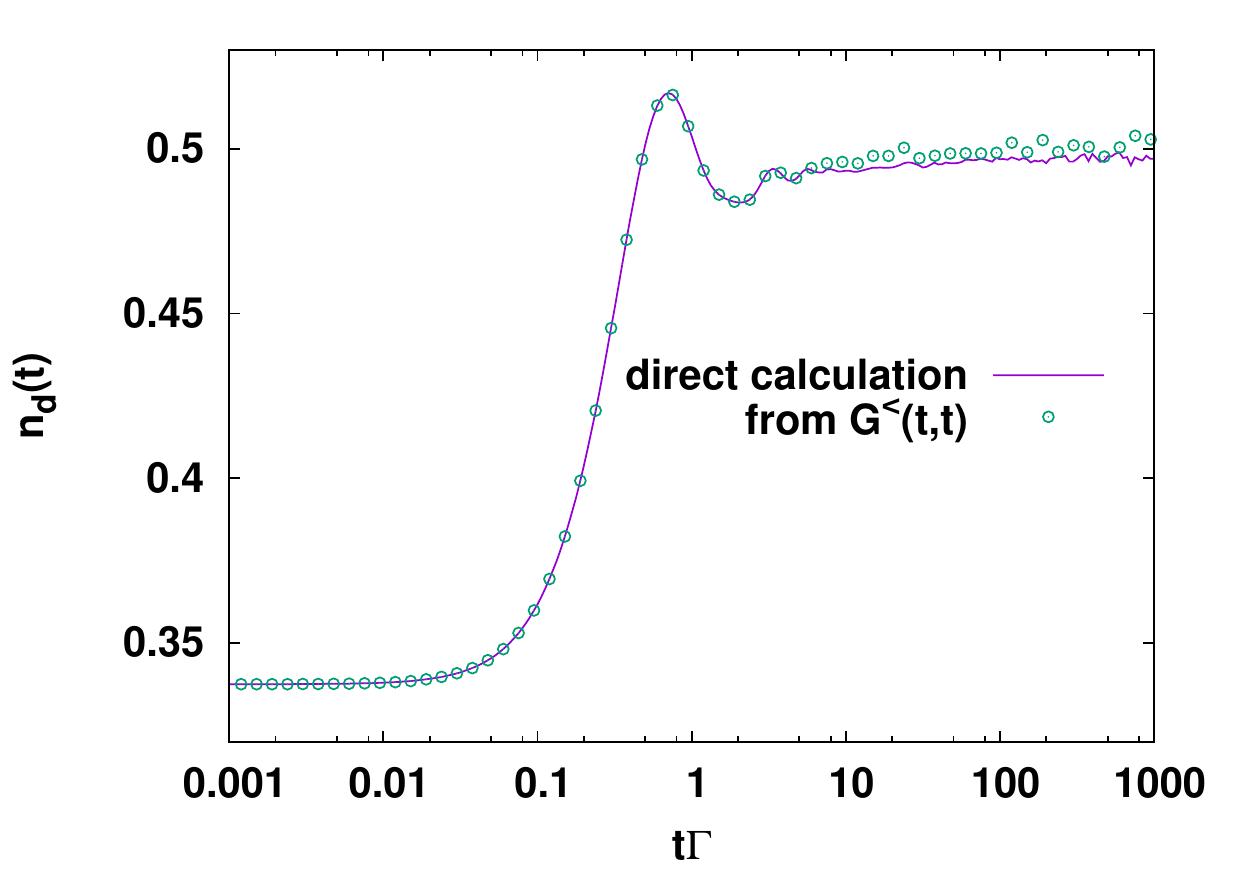}
  \caption{Time evolution of the occupation number (per spin) calculated directly  
    as a thermodynamic observable \cite{S_Nghiem2014a} and calculated from  
    the equal time lesser Green function. The system is driven from an  
    asymmetric mixed valent to a symmetric Kondo regime as in quench  
    (b) of the main text.}\label{fig:compare_static_dynamic}
\end{figure}

\subsection{Retarded Green function: explicit dependence on times}
\label{subsec:gf-real-time}
From Eq.~(\ref{eq:gf-new-positive}), we can directly evaluate the
dependence of the retarded Green function on its two time
arguments. This is shown for the imaginary and real parts in
Figs.~\ref{fig:G_2time}(a)-\ref{fig:G_2time}(b) versus the time difference $t'>0$ and
time $t>0$. At equal times we see from Fig.~\ref{fig:G_2time}(a) that $-{\rm
  Im [G(t,t))]}=1$ for all $t>0$, recovering the canonical
anticommutation relation for fermions, ande hence the spectral
sum rule for $t>0$. Non-equilibrium DMFT \cite{S_Freericks2006,S_Schmidt2002,S_Gramsch2013,S_Aoki2014}
requires impurity Green functions in real time, and the ability to
calculate these within TDNRG, which we here demonstrated, is a useful
first step for future applications to the former. 
\begin{figure}[t]
    \includegraphics[width=1.0\textwidth]{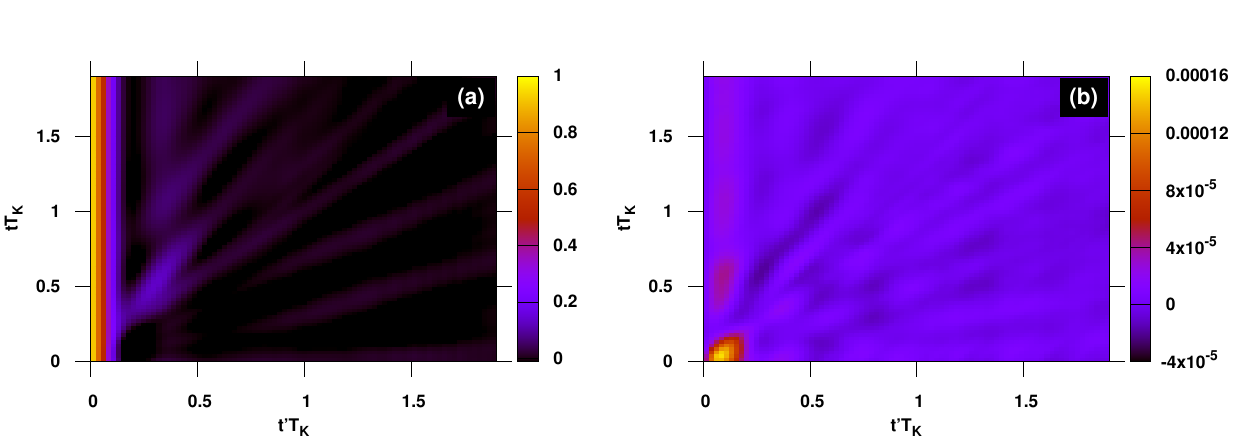}
  \caption{(a) Imaginary part, $-{\rm Im [G(t'+t,t))]}$, of the retarded
    Green function  {\it vs} the time difference $t'>0$ and the
    time $t>0$ for a quench in which $\varepsilon_{i}\gg \Gamma$ and
    $U_{i}=6\Gamma$ (corresponding to an initially empty orbital) and
    $\varepsilon_{f}=-3\Gamma$ with $U_{f}=6\Gamma$ (such that the
    final state is Kondo correlated with $T_{\rm K}/\Gamma\approx
    0.2$). This may be compared similar
    results from continuous time Quantum Monte Carlo (Fig.~3 of
    Ref.~\onlinecite{S_Cohen2014a}). TDNRG parameters: discretization
    parameter $\Lambda=4$, $z$ averaging with $N_{z}=64$, energy
    cut-off $E_{\rm cut}=24$. (b) Real part, ${\rm Re [G(t'+t,t))]}$, of the same retarded Green
    function  {\it vs.} the time difference $t'>0$ and the
    time $t>0$}
\label{fig:G_2time}
\end{figure}
\clearpage
\end{document}